\documentclass[a4paper,runningheads]{llncs}
\pdfoutput=1
\usepackage[utf8]{inputenc}
\usepackage[T1]{fontenc}
\usepackage[hidelinks]{hyperref}
\usepackage{xcolor}

\usepackage{orcidlink}\renewcommand{\orcidID}[1]{\orcidlink{#1}}

\usepackage[title]{appendix}
\usepackage{xspace}
\usepackage{array}
\usepackage{booktabs}
\usepackage{fancyvrb}
\usepackage{amssymb}
\usepackage{amsfonts}
\usepackage{amsmath}
\usepackage{bm}
\usepackage{tikz}
\usetikzlibrary{matrix,snakes,arrows,shapes,positioning,arrows.meta,shapes.geometric}
\newcommand{\y}{\textbullet}
\newcommand{\pl}[1]{\texttt{#1}}  %
\newcommand{\plv}[1]{\textrm{\textit{#1}}} %
\newcommand{\al}[1]{\textbf{#1}} %

\newcommand{\imp}{\rightarrow}

\newcommand{\la}{\langle}
\newcommand{\ra}{\rangle}
\newcommand{\eqdef}{\; 
\raisebox{-0.1ex}[0mm]{$ \stackrel{\raisebox{-0.2ex}{\tiny 
      \textnormal{def}}}{=} $}\; }
\newcommand{\defname}[1]{\emph{#1}}
\newcommand{\name}[1]{\emph{#1}}
\newcommand{\f}[1]{\mathsf{#1}}
\newcommand{\g}[1]{\mathit{#1}}

\renewcommand{\P}{\f{P}}
\renewcommand{\i}{\f{i}}
\newcommand{\D}{\f{D}}
\newcommand{\Det}{\text{\textit{Det}}\xspace}

\newcommand{\Syll}{\text{\textit{Syll}}\xspace}

\newcommand{\Lukasiewicz}{{\L}u\-ka\-sie\-wicz\xspace}
\newcommand{\fa}{\f{a}}
\newcommand{\fb}{\f{b}}
\newcommand{\fc}{\f{c}}

\newcommand{\Vampire}{\textsf{Vampire}\xspace}
\newcommand{\OTTER}{\textsf{OTTER}\xspace}
\newcommand{\SGCD}{\textsf{SGCD}\xspace}
\newcommand{\EProver}{\textsf{E}\xspace}
\newcommand{\ProverN}{\textsf{Prover9}\xspace}
\newcommand{\leanCoP}{\textsf{leanCoP}\xspace}
\newcommand{\CCS}{\textsf{CCS}\xspace}
\newcommand{\CCSV}{\textsf{CCS-Vanilla}\xspace}
\newcommand{\CMProver}{\textsf{CMProver}\xspace}
\newcommand{\SETHEO}{\textsf{SETHEO}\xspace}
\newcommand{\PTTP}{\textsf{PTTP}\xspace}
\newcommand{\PIE}{\textsf{PIE}\xspace}
\newcommand{\CDTools}{\textsf{CD Tools}\xspace}

\newcolumntype{L}[1]{>{\raggedright\let\newline\\\arraybackslash\hspace{0pt}}p{#1}}

\newcommand{\txtcomment}[2]%
{{\color{blue}{\textbf{#1:} #2}}}

\definecolor{revcolor}{rgb}{0.8,0,0}

\definecolor{revdonecolor}{rgb}{0,0.4,0}

\newcommand{\xtoprule}{}
\newcommand{\xbottomrule}{}

\newcommand{\supplementarynote}{Supplementary material is provided in the
   appendix}
 
\newcommand{\appref}[1]{App.~\ref{#1}}

\begin{document}

\title{Lemmas: Generation, Selection, Application\thanks{ Funded by
    the Deutsche Forschungsgemeinschaft (DFG, German Research
    Foundation) -- Project-ID~457292495, by the North-German
    Supercomputing Alliance (HLRN), by the ERC grant CoG ARTIST
    101002685, by the Hungarian National Excellence Grant
    2018-1.2.1-NKP-00008 and the Hungarian Artificial Intelligence
    National Laboratory Program (RRF-2.3.1-21-2022-00004).}}
\author{Michael Rawson~\inst{1}\orcidID{0000-0001-7834-1567} \and
  Christoph Wernhard~\inst{2}\orcidID{0000-0002-0438-8829} \and Zsolt
  Zombori~\inst{3}\orcidID{0000-0001-8622-5304} \and\\ Wolfgang
  Bibel~\inst{4}\orcidID{0000-0003-3892-0171} }
\authorrunning{M.~Rawson et al.}  \institute{ TU Wien,
  Austria~\email{michael@rawsons.uk} \and University of Potsdam,
  Germany~\email{info@christophwernhard.com} \and Alfr\'ed R\'enyi
  Institute of Mathematics, Hungary~\email{zombori@renyi.hu}\and
  Technical University Darmstadt, Germany~\email{bibel@gmx.net}}
\maketitle
\begin{abstract}
Noting that lemmas are a key feature of mathematics, we engage in an
investigation of the role of lemmas in automated theorem proving. The paper
describes experiments with a combined system involving learning technology
that generates useful lemmas for automated theorem provers, demonstrating
improvement for several representative systems and solving a hard problem not
solved by any system for twenty years. By focusing on condensed detachment
problems we simplify the setting considerably, allowing us to get at the
essence of lemmas and their role in proof search.
\end{abstract}

\section{Introduction}
\label{sec:intro}
Mathematics is built in a carefully structured way, with many disciplines and
subdisciplines. These are characterized by concepts, definitions, axioms,
theorems, lemmas, and so forth. There is no doubt that this inherent structure
of mathematics is part of the discipline's long-lasting success.

Research into Automated Theorem Proving (ATP) to date has taken little notice of the
information provided by this structure. Even state-of-the-art ATP systems
ingest a conjecture together with pertinent definitions and axioms
in a way completely agnostic to their place in the mathematical structure.
A comparatively small but nevertheless important part of the structure of mathematics
is the identification and application of \emph{lemmas}. It is
this aspect which is the focus of the work presented here.

The purpose of lemmas in mathematics is at least threefold.
First, and perhaps most importantly, lemmas
support the search for proofs of assertions. If some lemma applies
to a given problem, a proof may be found more easily.
Second, it is often the case that a lemma may be applied more than once.
If this happens, its use will shorten the length of the overall proof since
the proof of the lemma need only be carried out once, not repeatedly
for every application.
Third, the structuring effect of proofs by
the use of lemmas is an important feature for human comprehension of proofs.
In our work we are motivated primarily by the first two of these three aspects.

These considerations give rise to the crucial question: how can we find useful
lemmas for proving a given problem?  Here we mean useful in the sense of the
two aforementioned aspects: lemmas should be applicable to the problem at
hand, preferably many times.  In full generality this is a difficult question
indeed, which will require much further research. In this first step we
restrict the question to a narrow range of problems, known in literature as
\emph{condensed detachment} (CD) problems \cite{meredith:notes:1963}.  Proofs
of CD problems can be represented in a simple and accessible form as
\emph{proof structure terms}, enabling structure enumeration to enhance proof
search and lemma maintenance, as well as feature extraction for learning.
Our investigation thus focuses on the question of how ATP performance may be
improved for CD problems by the generation and selection of useful lemmas
before search begins.

CD problems are of the form ``axiom(s) and \Det imply a goal'' where \Det
represents the well-known modus ponens rule, or \emph{condensed detachment}. They
have a single unary predicate. A typical application is the investigation of
an axiomatization of some propositional logic, whose connectives are then
represented by function symbols. In order to support this study
experimentally, we have built a combined system for dealing with these
problems. It features \SGCD~\cite{cw:sgcd} as prover and lemma generator along
with a learning module based on either an easily-interpreted linear model over
hand-engineered features, or a graph neural network supporting end-to-end
learning directly from lemmas.

Our work results in a number of inter-related particular contributions:
\begin{enumerate}
\item Incorporation of proof structure terms into ATP with Machine
  Learning (ML). Consideration of features of the proof structure
  terms, explicitly in linear-model ML or implicitly in a neural ML
  model. A novel ATP/ML dataflow that is centered around proof
  structure terms.

\item Experimentally validated general insights into the use of learned lemmas
  for provers of different paradigms, with different ways to incorporate
  lemmas, and based on two alternate ML models. At the same time pushing
  forward the state of the art on proving CD problems. Insights include: \SGCD
  is competitive with leading first-order provers; Learned lemmas
  significantly extend the set of problems provable by the leading first-order
  prover \Vampire; Provers without internal lemma maintenance, such as
  Connection Method (CM)
  \cite{bibel:atp:1982,bibel:deduction:1993,bibel:otten:2020} systems, are
  drastically improved; \Vampire and \SGCD are able to handle a few hundreds
  of supplied lemmas; Learning based on manual features and on automatic
  feature extraction perform similarly.

\item An automatic proof of the Meredith single axiom theorem
  \texttt{LCL073-1}, which has persisted in the TPTP rated 1.00 since 1997.
  The first and only system to succeed was \OTTER \cite{otter}, after
  intensive massaging by Wos~\cite{wos:meredith}. It was proven by \SGCD in a
  novel systematic way.
  
\item An implemented framework with the new techniques for generation,
  selection and application of lemmas.
\end{enumerate}

\subsubsection{Structure of the Paper.}
Section~\ref{sec:background} presents condensed detachment and its embedding into the CM by way of so-called \emph{D-terms},
as well as background material on lemmas and machine learning in ATP.
Section~\ref{sec:ml} introduces a method for generating and selecting useful lemmas
and presents experimental results with it,
leading up to the proof of \texttt{LCL073-1} in Sect.~\ref{sec:LCL073}.
We conclude with a summary and outlook for further work in this area
in Sect.~\ref{sec:conclusion}.

\supplementarynote.
All experiments are fully reproducible and the artifacts are available at
\url{https://github.com/zsoltzombori/lemma}, commit \texttt{df2faaa}. We use
\CDTools~\cite{cw:sgcd} and \PIE \cite{cw:pie:2016,cw:pie:2020}, implemented
in SWI-Prolog~\cite{swiprolog}, for reasoning tasks and PyTorch~\cite{pytorch}
for learning.

\section{Background and Related Work}
\label{sec:background}

In a very general sense, lemmas in ATP factorize duplication. This may be
between different proofs that make use of the same lemma, or within
a single proof where a lemma is used multiple times.
It may not even be a particular formula that is shared, but a \emph{pattern},
such as a \name{resonator}~\cite{wos:resonance:95}.
In the presence of machine learning, we may think of even more abstract entities
that are factorized: the \emph{principles} by which proofs are written,
repeated in different proofs or contexts.

Depending on the proving method, lemmas in ATP play different
roles.
Provers based on \emph{saturation}, typically resolution/superposition (RS) systems~\cite{saturation},
inherently operate by generating lemmas: a resolvent is itself a lemma derived from its parents.
Nevertheless, one may ask for more meaningful lemmas than the clauses of the proof.
This is addressed with \name{cut introduction}~\cite{woltzenlogel:cutintro:2010,cut:intro:2012,vienna:qcuts:2019},
which studies methods to obtain complex lemmas from resolution proofs.
Such lemmas provide insight about the high-level structure of proofs,
extract interesting concepts and support research into the
correspondence between natural mathematical notions and possible proof compressions.
Other approaches to interesting theorems or lemmas are described for example in~\cite{sutcliffe:2003:theorem:discovery,pudlak:lemmas:2006}.

Another question concerning lemmas and ATP systems is whether performance can
be improved by supplementing the input with lemmas.  This is particularly
applicable if lemmas are obtained with methods that are \emph{different} from
those of the prover. Otherwise, it may have obtained these by
itself.\footnote{We note here that in some cases systems \emph{cannot}
generate certain lemmas because of e.g. ordering restrictions.}  As we will
see, leading ATP systems such as \Vampire and \EProver \cite{eprover} can
indeed be improved in this way.  Different \emph{methods} does not necessarily
mean different \emph{systems}: it is possible to use different configurations
of the same system for lemma generation and proving, as well as for
intermediate operations.  This was the workflow used by Larry Wos to prove the
challenge problem \texttt{LCL073-1} with \OTTER \cite{wos:meredith}. Our \SGCD
system also supports this, which played a major role in its ability to prove
the aforementioned challenge problem.

Lemmas play a quite different role for a family of provers which we
call \name{CM-CT} for \name{Connection Method/Clausal Tableaux},
exemplified by \PTTP~\cite{pttp}, \SETHEO \cite{setheo:92}, and
\leanCoP \cite{otten:2003:leancopinabstract,leancop}. Underlying
conceptual models are model elimination \cite{loveland:1978}, clausal
tableaux \cite{letz:habil} and the CM. They enumerate proof structures
while propagating variable bindings initialized by the goal through
unification, and hence proceed in an inherently goal-driven way. While
they are good at problems that benefit from goal direction, in general
they are much weaker than RS provers and have not been among the top
provers at \name{CASC} for about two decades. This is attributed to
the fact that they do not re-use the proof of one subgoal as the
solution of another: they do not use lemmas \emph{internally}.

The lack of lemmas was identified early as a weakness of CM-CT
\cite{eder:cs:1989}, so there have been various proposed
remedies~\cite{eder:cs:1989,astrachan:stickel:caching:1992,stickel:upsidedown:1994,schumann:delta:1994,letz:cut:1994,fuchs:lemmas:ijcai:1999,leancop,etableau}.
Despite some insight and success, this did not yet elevate CM-CT to the level
of the best RS systems.
Nevertheless, the expectation remains that CM-CT provers would benefit from
supplying lemmas as additional input. Hence, we included two CM-CT systems in
our experiments, \leanCoP and
\CMProver\cite{cw:mathlib:97,cw:pie:2016,cw:pie:2020} and show that the
expectation is greatly confirmed.  Two other systems considered here, \SGCD
and \CCS~\cite{cw:ccs}, can be viewed as CM-CT systems extended to support
specific forms of lemma generation and application.

Lemmas can be maintained within the prover as an inherent part of the
method, as in saturation.
They may also be created and applied by different systems,
or different instances of the same system~\cite{discount,vampire-coop}.
Larry Wos calls this \name{lemma adjunction}~\cite{wos:lemma:inclusion:adjunction}.
Lemmas created by one system are passed to a second system in two principal
ways.
First, they can be passed as \emph{additional axioms}, in the hope
that the second system finds a shorter proof in the wider
but shallower search space. Second, external lemmas can be used to
\emph{replace search}. The second system then starts with the given
lemmas as if they were the cached result of its previous computation.
Moreover, the provided lemmas can be restricted in advance by
heuristic methods, such as by a machine-learned model. \SGCD
supports this \name{replacing} lemma incorporation. The basic distinction
between augmenting and replacing search with lemmas was already
observed by Owen L. Astrachan and Mark E. Stickel
\cite{astrachan:stickel:caching:1992} in the context of improving
CM-CT provers.

\subsection{Machine Learning for ATP}
\label{sec:ml:atp}

The past decade has seen numerous attempts to leverage machine
learning in the automated theorem proving effort. Early systems mostly
focused on premise selection,
e.g. \cite{malarea,deepmath,formulanet}, aiming to reduce the number
of axioms supplied as input to the prover,
or on selection of heuristics, e.g. \cite{bridge-holden-paulson-2014}.
Other works provide
internal guidance directly at the level of inferences during search,
e.g. \cite{deep_guidance,enigma,tactictoe,rlcop,plcop,lazycop:2021,holden:fnt-2021}. The
emergence of generative language models has also led to some
initial attempts at directly generating next proof steps,
e.g. \cite{piotrowski:rnn,conjecturing_dataset,openai_transformer_atp}, moving
the emphasis away from search.

In contrast to these lines of work, our focus is on learning the
utility of lemmas. Close to our aims is
\cite{KALISZYK2015109,DBLP:conf/frocos/KaliszykUV15}, trying to
identify globally useful lemmas in a collection of millions of proofs
in HOL Light. Besides differences in the formal system, what
distinguishes our work is that we learn a much more focused model: we
put great emphasis on evaluating lemmas in the context of a particular
goal and axiom set; in fact, our entire system was designed around the
question whether a given lemma is moving the goal closer to the
axioms. We argue that the D-term representation of all involved components
(goal, lemma, axioms, proof) makes our framework particularly suitable
for the lemma selection task.

We employ an iterative improvement approach first used in
MaLARea~\cite{malarea}: in each iteration, we run proof search guided by a
learned model, extract training data from proving attempts, and fit a new
model to the new data. These steps can be repeated profitably until
performance saturates.

\subsection{Condensed Detachment: Proofs as Terms}

\label{sec:cd:atp}

\name{Condensed detachment (CD)} was developed in the mid-1950s by Carew A.
Meredith as an evolution of \name{substitution and detachment}
\cite{prior:logicians:1956,lemmon:meredith:purestrict:1957,prior:formal:logic:1962,meredith:memoriam:1977}.
Reasoning steps are by \name{detachment}, or modus ponens, under implicit
substitution by most general unifiers. Its primary application is the
investigation of axiomatizations of propositional logics at a first-order
meta-level.
CD also provides a technical approach to the Curry-Howard correspondence,
``formulas as types''~\cite{hindley:meredith:cd:1990,hindley:book:1997} and
is considered in witness theory \cite{rezus:2020:witness}.
Many early successes in ATP were on CD
problems~\cite{mccune:wos:cd:1992,ulrich:legacy:2001}, but success was also
found in the reverse direction. Refinements of the \OTTER prover
in the 1990s, some of which have found their ways into modern RS provers, were
originally conceived and explored in the setting of CD
\cite{wos:contributes:1990,wos:bledsoe:91,mccune:wos:cd:1992,wos:resonance:95,wos:combining:96,veroff:shortest:2001,fitelson:missing:2001,wos:meredith}.

From a first-order ATP perspective, a CD problem consists of \name{axioms},
i.e. positive unit clauses; a \name{goal theorem}, i.e. a single negative
ground unit clause representing a universally-quantified atomic goal theorem
after Skolemization; and the following ternary Horn clause that models
detachment.
\[\Det\; \eqdef\; \P(\i(x,y)) \land \P(x) \imp \P(y).\]
The premises of \Det are called the \name{major} and \name{minor} premise,
respectively. All atoms in the problem have the same predicate $\P$, which is
unary and stands for something like \name{provable}. The formulas of the
investigated propositional logic are expressed as terms, where the binary
function symbol $\i$ stands for \name{implies}.

CD may be seen as an \emph{inference rule}. From an ATP perspective, a
\name{CD inference step} can be described as a hyperresolution
from \Det and two positive unit clauses to a third positive unit clause.
A \defname{CD proof} is a
proof of a CD problem constructed with the CD inference rule. CD proofs can be
contrasted with other types of proof, such as a proof with binary resolution
steps yielding non-unit clauses. \ProverN \cite{prover9} chooses positive
hyperresolution by default as its only inference rule for CD problems and thus
produces CD proofs for these.

It is, however, another aspect of CD that makes it of particular interest for
developing new ATP methods, which only recently came to our attention in the
ATP context~\cite{cwwb:lukas:2021}: the structure of CD proofs
can be represented in a very simple and convenient way as full binary trees,
or as terms. In ATP we find this aspect in the CM, where the proof structure
as a whole is in focus, in contrast to extending a set of formulas by
deduction~\cite{bonacina:taxonomy}. This view of CD is made precise and
elaborated upon in \cite{cwwb:article}, on which the subsequent
informal presentation is based. We call the structure
representations of CD proofs \defname{D-terms}. A D-term is a term recursively
built from numeral constants and the binary function symbol~$\D$ whose
arguments are D-terms. In other words, it is a full binary tree where the leaf
nodes are labeled with constants. Four examples of D-terms are
\[1,\;\;\; 2,\;\;\; \D(1,1),\;\;\; \D(\D(2,1),\D(1,\D(2,1))).\]
A D-term represents the structure of a proof. A proof in full is represented
by a D-term together with a mapping of constant D-terms to axioms. Conversion
between CD proofs and D-terms is straightforward: the use of an axiom corresponds to a constant
D-term, while an inference step corresponds to a D-term $\D(d_1,d_2)$ where $d_1$
is the D-term that proves the major premise and $d_2$ the minor.

Through first-order unification, constrained by axioms for the leaf nodes and
the requirements of \Det for inner nodes, it is possible to obtain a most
general formula proven by a D-term~\cite{cwwb:article}. We call
it the \defname{most general theorem} (MGT) of the D-term with respect to the
axioms, unique up to renaming of variables.
For a given axiom map, not all D-terms necessarily have an MGT: if unification fails,
we say the D-term has no MGT. It is also possible that different
D-terms have the same MGT, or that the MGT of one is subsumed by the MGT of
another.
A D-term is a proof of the problem if its MGT subsumes the goal theorem.

As an example, let the constant D-term~$1$ be mapped to $\P(\i(x,\i(x,x)))$,
known as \name{Mingle}~\cite{ulrich:legacy:2001}. Then, the MGT of the D-term
$1$ is just this axiom. The MGT of the D-term $\D(1,1)$ is
$\P(\i(x,\i(x,x)),\i(x,\i(x,x)))$, that is, after renaming of variables,
$\P(y)\sigma$ where $\sigma$ is the most general unifier of the set of pairs
$\{\{\P(\i(x,y)),\, \P(\i(x',\i(x',x')))\},\;
 \{\P(x),\, \P(\i(x'',\i(x'',x'')))\}\}$.

D-terms, as full binary trees, facilitate characterizing and investigating
structural properties of proofs. While, for a variety of reasons, it is far
from obvious how to measure the size of proofs obtained from ATP systems in
general, for D-terms there are at least three straightforward size measures:
\begin{itemize}
\item The \defname{tree size} of a D-term is the number of its inner nodes.
\item The \defname{height} of a D-term is the length of the longest root-leaf path.
\item
  The \defname{compacted size} of a D-term is the number of distinct compound
  subterms, or, in other words, the number of inner nodes of its minimal
  DAG.
  
\end{itemize}
Alternative names in the literature are \name{length} for compacted
size, \name{level} for height and
\name{CDcount}~\cite{veroff:shortest:2001} for tree size.  The D-term
$\D(\D(1, \D(1, 1)), \D(\D(1, 1), 1))$, for example, has tree size~5,
compacted size~4 and height~3.
\name{Factor equations} provide a compact way of writing D-terms: distinct subproofs with
multiple incoming edges in the DAG receive numeric labels, by which they are
referenced. The D-term $\D(\D(1, 1), \D(\D(1, \D(1, 1)), \D(1, \D(1, 1))))$,
for example, can be written as
$2 =  \D(1, 1),\; 3 = \D(1, 2),\; 4 = \D(2, \D(3, 3))$.

CD problems have core characteristics of first-order ATP problems: first-order
variables, at least one binary function symbol and cyclic predicate
dependency. But they are restricted: positive unit clauses, one negative
ground clause, and one ternary Horn clause. Equality is not explicitly
considered. The generalization of CD to arbitrary Horn problems is, however,
not difficult \cite{cw:ccs}.

\subsection{Condensed Detachment for ATP and Lemmas}
\label{sec:cd:atp:lemmas}

From an ATP point of view, D-terms provide access to proofs as a whole.  This
exposes properties of proofs that are not merely local to
an inference step, but spread across the whole proof. It
suggests a shift in the role of the calculus from providing a recipe for
building the structure towards an inductive structure
\emph{specification}. Moreover, D-terms as objects provide insight into
\emph{all} proofs: for example, growth rates of the number of binary trees for
tree size, height and compacted size are well-known with entries in \name{The
  On-Line Encyclopedia of Integer Sequences}~\cite{oeis} and provide upper
bounds for the number of proofs \cite{cwwb:article}. A practical consequence
for ATP is the justification of proof structure enumeration techniques where
each structure appears at most once.
  
CD proofs suggest and allow for a specific form of lemmas, which we call
\name{unit} or \name{subtree} lemmas, reflecting two views on them.
As formulas, they are positive unit clauses, which can be re-used in different CD inference
steps. In the structural view, they are subterms, or subtrees, of the overall
D-term. If they occur multiply there, they are factored in the minimal
DAG of the overall D-term. The views are linked in that the formula of a
lemma is the MGT of its D-term.
The \emph{compacted size} measure specified above takes into account
the compression achievable by unit/subtree lemmas.
From the perspective of proof structure compression methods,
unit/subtree lemmas have the property that the compression target is unique,
because each tree is represented by a unique minimal DAG.
CM-CT provers do not support such lemmas, which is the main reason for
their notorious weakness on CD problems.

\subsection{SGCD ---  Structure Generating Theorem Proving}
\label{sec:sgcd}

\SGCD \name{(Structure Generating Theorem Proving for Condensed
  Detachment)}
\cite{cw:sgcd} is the central system used in our experiments as prover as well
as lemma generator. It realizes an approach to first-order theorem proving
combining techniques known from the CM and RS that was not fully recognized
before. It generalizes (for CD problems) bottom-up preprocessing for and with
CM-CT provers \cite{schumann:delta:1994} and hypertableaux
\cite{hypertableaux}.
\SGCD works by enumeration of proof structures together with
unification of associated formulas, which is also the core method of the CM-CT
provers. Structures for which unification fails are excluded.
Each structure appears at most once in the enumeration.

Let the proof structures be D-terms. Partition the set of all D-terms
according to some \name{level} such that those in a lower level are strict
subterms of those in a higher level. Tree size or height are examples of such
a level. Let
\begin{center}
  \pl{enum\_dterm\_mgt\_pairs(\plv{+Level}, \plv{?DTerm}, \plv{?Formula})}
\end{center}
be a Prolog\footnote{Prolog serves here as a suitable specification language.} predicate enumerating D-terms and corresponding MGTs at a certain level,
with respect to given axioms that do not explicitly
appear as parameter.
We say that the predicate generates these pairs
in an \defname{axiom-driven} way.
If the predicate is invoked with the formula argument instantiated by a ground
formula, it enumerates D-terms that prove the formula at the specified level.
The predicate is then used \defname{goal-driven}, like a CM-CT prover.
Invoking it for increasing level values realizes iterative deepening.
There are further instantiation possibilities: if only the D-term is
instantiated and the level is that of the D-term, its MGT is computed. If both
D-term and formula are instantiated, the predicate acts as verifier.

The implementation includes several \name{generators},
concrete variants of the \pl{enum\_dterm\_mgt\_pairs} predicate for specific
level characterizations. \SGCD maintains a cache of $\la \plv{level},
\plv{D-term}, \plv{formula} \ra$ triples
used to
obtain solutions for subproblems in levels below the calling level.
This cache is highly configurable. In particular, the number of entries can be
limited, where only the best triples according to specified criteria are kept.
Typical criteria are height or size of the formula, a heuristic shared with RS provers.
Subsumed entries can be deleted, another feature in common with RS.
Novel criteria are also supported, some of which relate the formula to the goal.
Most criteria are based on the formula component of the triples, the MGT.
Due to rigid variables~\cite{handbook:ar:haehnle}, MGTs are not usually available in
CM-CT provers~\cite{cwwb:article} and cannot be used as a basis for heuristics.

When lemmas are provided to \SGCD, they are used to initialize the cache,
replacing search at levels lower than the calling level.\footnote{Replacement can be
subject to heuristic restrictions.}
\SGCD further maintains a set of \name{abandoned} $\la \plv{level},
\plv{D-term}, \plv{formula} \ra$ triples, those that are generated but do not
qualify for entering the cache or were removed from the cache. These are kept
as a source for heuristic evaluation of other triples and for lemma generation.

For theorem proving, \SGCD proceeds as shown in Fig.~\ref{fig-sgcd-pseudocode}.
Input parameter $g$ is the goal formula, while parameters $\g{maxLevel}$ and $\g{preAddMaxLevel}$ are configurable.
\pl{enum\_dterm\_mgt\_pairs} represents a particular
generator that is also configurable. It enumerates argument
bindings nondeterministically: if it succeeds in the inner loop,
an exception returns the D-term $d$. $C$ is the
cache. The procedure \pl{merge\_news\_into\_cache($N,C$)} merges newly
generated $\la \plv{level}, \plv{D-term}, \plv{formula}\ra$ triples~$N$ into
the cache~$C$.
If $\g{maxLevel}$ is configured as 0, the method proceeds in purely
goal-driven mode with the inner loop performing iterative deepening on the
level~$m$. Similarity to CM-CT provers can be shown empirically by
comparing the sets of solved TPTP problems \cite{cw:sgcd}.
Generally successful configurations of $\g{preAddMaxLevel}$ typically have
values 0--3.

\begin{figure}[t]
  \centering

\begin{tabular}{l}
  $C := \emptyset$;\\
  \al{for} $\g{l} := 0$ \al{to} $\g{maxLevel}$ \al{do}\\
  \hspace{2em} \al{for} $\g{m} := \g{l}$ \al{to} $\g{l}+\g{preAddMaxLevel}$ \al{do}\\
  \hspace{4em}\pl{enum\_dterm\_mgt\_pairs($m,d,g$)};\\
  \hspace{4em}\al{throw} \pl{proof\_found($d$)}\\
  \hspace{2em} $N := \{\la l,d,f \ra \mid\; $\pl{enum\_dterm\_mgt\_pairs($l,d,f$)}$\}$;\\
  \hspace{2em} \al{if} $N = \emptyset$ \al{then throw} \pl{exhausted};\\
  \hspace{2em} $C :=$ \pl{merge\_news\_into\_cache($N,C$)}
\end{tabular}
\caption{The nested loops of the \SGCD theorem proving method.}
\label{fig-sgcd-pseudocode}
\end{figure}

\section{Improving a Prover via Learned Lemma Selection}
\label{sec:ml}

We employ machine learning to identify lemmas that can enhance proof
search. Unlike the standard supervised scenario in which we learn from
some training problems and evaluate performance on separate test
problems, we take a reinforcement learning approach of self-improvement
that has already been successfully applied in several
theorem proving projects since \cite{malarea}. In this approach, we
perform proof search with a \emph{base prover} on our entire problem
set and learn from the proof attempts.\footnote{We currently only
learn from successful proof attempts and sketch an extension to
learning from failure.} The learning-assisted prover is evaluated
again in the problem set to see if it can find more or different
problems. If there is improvement, the process can be repeated until
performance saturates.
In a bit more detail, our system has the following components.

\begin{enumerate}
  \item {\bf Base Prover}: Performs proof search and its main role is
    to provide training data to the utility model. 
  \item {\bf Utility Model}: The model takes $\la$\textit{conjecture},
    \textit{lemma}, \textit{axioms}$\ra$ triples and outputs a utility score,
    i.e., some measure of how useful the lemma is for proving the conjecture
    from the axioms. The utility model is trained from the D-terms emitted by
    the base prover.
  \item {\bf Lemma Generator}: Produces a large set of
    candidate lemmas for each problem separately. All candidates are
    derivable from the axioms.
  \item {\bf Evaluated Prover}: For each problem, we evaluate the
    candidate sets with the utility model and select the best
    ones. These lemmas are provided to the evaluated prover which
    performs proof search on the problem set. The evaluated prover can
    be identical to or different from the base prover.
\end{enumerate}

\subsubsection{Base Prover.}
Any prover that emits proofs as D-terms is suitable as a base prover. Given a
D-term proof tree $P$ of some formula $C$ from axiom set $As$, any connected
subgraph $S$ of $P$ can be considered as the proof of a lemma $L$. If $S$ is a
full tree, it proves a unit lemma, which is the formula associated with its
root. Otherwise, it proves a Horn clause, whose head is the root formula of
$S$ and whose body corresponds to the open leaves of $S$. We currently focus on
unit lemmas and leave more general subgraphs for future work. To approximate
the utility of lemma $L$ for proving $C$ from $As$, there are several
easy-to-compute logical candidates, such as the reduction in tree size, tree
height or compressed size. A more refined measure is obtained if we reprove
$C$ with the lemma $L$ added to the axioms $As$ and observe how the number of
inference steps changes.\footnote{The number of inferences is a measure
provided by the Prolog engine and is not identical to the number of steps in
the FOL calculus.} This is slower to compute, but takes into account the
particularities of the base prover, hence provides more focused guidance. In
our experiments, we find that the best performance is obtained by reproving
and then computing utility $U$ as the inference step reduction normalized into
$[-1, 1]$, where $-1$ means that the problem could not be solved within the
original inference limit and $1$ is assigned to the lemma that yields the
greatest speedup. We end up with tuples $\la C, As, L, U\ra$ to learn from.

\subsubsection{Utility Model Training.}
We experiment with gradient-descent optimization for two classes of functions:
linear models and graph neural networks (GNNs). Our linear model is based on
51 manually-identified features, some of them novel, described in
\appref{app:manualfeatures}. For each feature $f_i$ there is an associated
weight parameter $w_i$ to produce the final predicted utility
$$ U(\vec f; \vec w) = \sum_i f_i w_i$$ The second, more involved model is a
GNN. Describing this model is beyond the scope of this paper: see e.g.
\cite{gnn_intro} for a gentle introduction. What is crucial for our purposes
is that no manual feature extraction is involved: a specialized neural network
processes the D-terms of involved formulas directly and learns to extract
useful features during optimization. As input, the model is given a graph,
losslessly encoding D-terms of the lemma to be evaluated, the conjecture and
the axioms. The precise network architecture is provided in
\appref{app:implementation}.

\subsubsection{Candidate Lemma Generation.}
Candidate lemmas are generated separately for each problem via the
structure enumeration mechanism of \SGCD, as explained in
Fig.~\ref{fig-sgcd-pseudocode}. The goal $g$ is provided and
$\g{preAddMaxLevel}$ is set to 0, making \SGCD proceed axiom-driven,
generating lemmas level by level.  However, it does intersperse the
goal-driven inner loop, which is only trying to prove the goal on the
level directly above the last cached level.  \SGCD may terminate with
a proof, in which case further lemma generation is
pointless. Otherwise it terminates after $\g{maxLevel}$ is reached,
generation of new levels is exhausted, or a time limit is reached. We
then use the cache $C$ and the abandoned triples as the generated
output lemmas. Furthermore, there are many ways to configure \SGCD. We
obtained the best results generating by tree size and by PSP-level
(explained below), combined with known good heuristic restrictions.
In particular we restrict the size of the lemma formulas to the
maximum of the size of the axioms and the goal, multiplied by some
factor (usually 2--5).  We also restrict the number of elements in the
cache, typically to 1,000.  The lemmas are sorted by formula size
measures, smaller preferred, to determine which are retained in the
cache.

Proof structure generation by PSP-level is a novel technique introduced in
\cite{cw:sgcd,cwwb:article}, based on an observation by \Lukasiewicz and
Meredith. In a detachment step, often the D-term that proves one premise is a
subterm of the D-term that proves the other.  We turn this relationship into a
proof structure enumeration method: structures in level $n+1$ are D-terms
where one argument D-term is at level $n$ and the other argument is a subterm
of that D-term. The method is incomplete, but combines features of DAG
enumeration while being compatible with a simple global lemma maintenance as
realized with \SGCD's cache~\cite{cwwb:article}.

\subsubsection{Evaluated Prover.} For each problem, we evaluate the
candidate set with the utility model and select $k$ lemmas with the highest
predicted utility, where $k$ is a hyperparameter. The evaluated prover then
tries to solve the problems with the help of the selected lemmas. The lemmas
can either be treated as additional axioms --- applicable to any prover --- or
have a specialized treatment if the prover provides for it: in particular,
\SGCD and \CCSV use the lemmas to replace inner lemma
enumeration.\footnote{Before the obtained input lemmas are passed to a prover
we supplement them with the lemmas for all their subproofs, i.e. we close the
set of D-terms under the subterm relationship. This proved beneficial in
experiments (see, e.g., \appref{app:lemma:usage}). An alternative
would be to perform this closure on all generated lemmas before
selection.}
The evaluated prover can be any prover, since there is no
specialized requirement to handle lemmas as new axioms. If, however,
it is the base prover --- or any other system that emits proofs as
D-terms, then the learning procedure can be iterated as long as there
are new problems solved.

\subsection{Learning-Based Experiments}
\label{sec:experiment}

We experiment with a total of 312 CD problems, including all 196 pure
CD problems from TPTP~8.1.2~\cite{tptp},
enriched with single-axiom versions of all the problems to which a
technique by Tarski \cite{luk:tarski:aussagenkalkuel:1930}, as
specified by Rezuş \cite{rezus:tc:2016}, was applicable.  We test
several representative ATP systems, including state-of-the-art systems
for both general first-order reasoning and for CD problems.

\begin{table}[t]
  \centering
  \renewcommand{\arraystretch}{0.85}
\caption{Features of the considered provers: whether their proofs are
  available as D-terms (possibly after some conversion), whether they were used
  with \emph{replacing} lemma incorporation (Sect.~\ref{sec:background}),
  whether they operate goal-driven, and the underlying method.}
\label{tab-provers}
  \begin{tabular}{lcccccccc}
    & \SGCD & \ProverN & \CMProver & \leanCoP & \CCSV & \Vampire & \EProver\\
    \cmidrule(r{0.2em}l{0.2em}){2-2}\cmidrule(r{0.2em}l{0.2em}){3-3}\cmidrule(r{0.2em}l{0.2em}){4-4}\cmidrule(r{0.2em}l{0.2em}){5-5}
    \cmidrule(r{0.2em}l{0.2em}){6-6}\cmidrule(r{0.2em}l{0.2em}){7-7}\cmidrule(r{0.2em}l{0.2em}){8-8}
  D-terms & \y & \y & \y & $-$ & \y & $-$ & $-$\\
  Replacing lemmas & \y & $-$ & $-$ & $-$ & \y & $-$ & $-$\\
  Goal-driven & \y/$-$ & $-$ & \y & \y & \y & $-$ & $-$\\
  CM-CT & $-$ & $-$ & \y & \y & $-$ & $-$ & $-$\\
  RS & $-$ & \y & $-$ & $-$ & $-$ & \y & \y\\
  \end{tabular}
\end{table}

Table~\ref{tab-provers} gives an overview of the considered provers.
\CCSV is \CCS~\cite{cw:ccs} in a restricted configuration
to find only those CD proofs with minimal compacted size, identifying
problems that can clearly be solved with exhaustive search. It operates
goal-driven, like the CM-CT provers, but by enumerating DAGs instead of trees
through a local lemma maintenance mechanism. \Vampire and \EProver
represent the state of the art of first-order
ATP. Provers that produce D-terms as proofs (\SGCD, \ProverN,
\CMProver, \CCS) can serve as base provers. We always rely on \SGCD
for lemma candidate generation.
All provers are recent public versions: \Vampire 4.5.1,
\EProver~2.6, \leanCoP 2.1. We provide results in terms of \emph{time} limits, although for
the Prolog provers \SGCD, \CMProver and \CCSV we used
a roughly-equivalent inference limit to avoid fluctuations due to server
workload.

\subsubsection{Improving the Base Prover.}  In our first experiment,
we evaluate base provers after learning from their own proof
attempts. The provers are given $k=200$ best lemmas
according to the linear utility model.
Table~\ref{tab:baseprovers}\footnote{Further
visualizations of our experiments are provided in \appref{app:barchart}.}
shows problems solved by four base provers without lemmas (\name{Base}
case) and with two iterations of learning. The \name{Total} row gives
the number of theorems proved by any of the three iterations shown.
The stronger the base model, the harder it is to improve. \CMProver
and \CCSV are purely goal-driven and benefit greatly, reaching over
37\% improvement for larger time limits. \SGCD and \ProverN improve
over 5\% for shorter time limits, but this effect gradually vanishes
as the time limit is increased.

\begin{table}[htb]
  \caption{Number of problems solved over 2~iterations of
    training a linear model.}
  \label{tab:baseprovers}
  \centering
  \begin{tabular}{ l @{\hspace{5pt}} r r r r @{\hspace{5pt}} r r r r @{\hspace{5pt}} r r r r @{\hspace{5pt}} r r r r}
    \xtoprule
    & \multicolumn{4}{c}{\SGCD} & \multicolumn{4}{c}{\ProverN} & \multicolumn{4}{c}{\CMProver} & \multicolumn{4}{c}{\CCSV} \\
    Time & 50s & 100s & 500s & 30m & 50s & 100s & 500s & 30m & 50s & 100s & 500s & 30m & 50s & 100s & 500s & 30m \\
    \cmidrule(rl){2-5}\cmidrule(rl){6-9}\cmidrule(rl){10-13}\cmidrule(rl){14-17}
    Base   & 266 & 275 & 285 & 285 & 240 & 252 & 259 & 262 & 82 & 85 & 94 & 103 & 81 & 88 & 99 & 105 \\
    Iter 1 & 280 & 282 & 284 & 281 & 250 & 254 & 262 & 257 & 83 & 93 & 105 & 121 & 96 & 101 & 117 & 130 \\
    Iter 2 & 281 & 283 & 281 & 283 & 247 & 247 & 267 & 265 & 79 & 98 & 95 & 126 & 96 & 97 & 120 & 128 \\
    \cmidrule(rl){2-5}\cmidrule(rl){6-9}\cmidrule(rl){10-13}\cmidrule(rl){14-17}
    Total  & 282 & 284 & 286 & 286 & 253 & 258 & 269 & 267 & 91 & 105 & 112 & 141 & 106 & 105 & 133 & 145\\
    \xbottomrule
  \end{tabular}
\end{table}

An analysis, provided in \appref{app:lemma:usage}, reveals that in the
proofs not found during lemma generation and found by \SGCD after the
provision of lemmas, 63--96\% of the distinct subterms originate from the
lemmas, i.e., a substantial portion of the proofs are built up from the
provided lemmas.

\subsubsection{Learned Lemmas to Enhance other Provers.}  Next, we
fix \SGCD as base prover and evaluate other provers, namely \Vampire,
\EProver, \ProverN and \leanCoP.  Again, the provers are given $k=200$
best lemmas according to the linear utility model.
Table~\ref{tab:otherprovers} shows the greatest boost is for the
purely goal-driven \leanCoP, where there is over 40\% improvement for
all time limits. Second is \Vampire with 8--15\% improvement,
followed by \ProverN and \EProver with around 3\% improvement.
Interestingly, \EProver does not solve more problems with the lemmas,
but it solves different ones, hence the improvement. These results
suggest a great deal of transferability of the benefits of the lemma
selector.

\begin{table}[htb]
  \caption{Number of problems solved by \Vampire (casc), \EProver
    (autoschedule), \ProverN and \leanCoP without and with additional lemmas
    using various time limits.
  }
  \label{tab:otherprovers}
  \centering
  \setlength{\tabcolsep}{1pt}
  \begin{tabular}{ l @{\hspace{5pt}} r r r r @{\hspace{5pt}} r r r r
      @{\hspace{5pt}} r r r r @{\hspace{5pt}} r r r r }
    \xtoprule
    & \multicolumn{4}{c}{\Vampire} & \multicolumn{4}{c}{\EProver} & \multicolumn{4}{c}{\ProverN} & \multicolumn{4}{c}{\leanCoP} \\
    Time & 50s & 100s & 500s & 30m & 50s & 100s & 500s & 30m & 50s & 100s & 500s & 30m & 50s & 100s & 500s & 30m \\
    \cmidrule(rl){2-5}\cmidrule(rl){6-9}\cmidrule(rl){10-13}\cmidrule(rl){14-17}
    Base & 221 & 224 & 252 & 263 & 253 & 264 & 275 & 281 & 236 & 244 & 257 & 260 & 70 & 71 & 77 & 77 \\
    Lemmas & 249 & 257 & 274 & 283 & 256 & 266 & 275 & 275 & 246 & 250 & 261 & 269 & 100 & 103 & 111 & 113 \\
    \cmidrule(rl){2-5}\cmidrule(rl){6-9}\cmidrule(rl){10-13}\cmidrule(rl){14-17}
    Total & 249 & 257 & 276 & 284 & 269 & 276 & 287 & 286 & 248 & 252 & 264 & 269 & 100 & 103 & 111 & 113\\
    \xbottomrule
  \end{tabular}
\end{table}

\subsubsection{Changing the Number of Lemmas Added.}  Adding lemmas
has potential to shorten proofs, but it also widens the search space,
so it is not obvious how many lemmas are beneficial. In the next
experiment, we again fix \SGCD as base prover and evaluate \SGCD and
\Vampire with different number of lemmas
selected. Table~\ref{tab:lemmacount} shows that as little as 25 added
lemmas yield substantial improvement, 7\% for \Vampire and 4\% for
\SGCD, and performance does not drop as we add more lemmas: even at
500 we see no negative effect of the expanded search space.

\begin{table}[htb]
  \caption{Number of problems solved by \Vampire (casc) and \SGCD as we alter
    the number $k$ of supplemented lemmas. We use a time limit of 100s.}
  \label{tab:lemmacount}
  \centering
  \begin{tabular}{ l @{\hspace{5pt}} r r r r r r @{\hspace{5pt}} r r r r r r }
    \xtoprule
    & \multicolumn{6}{c}{\Vampire} & \multicolumn{6}{c}{\SGCD} \\
    Lemma count & 10 & 25 & 50 & 100 & 200 & 500 & 10 & 25 & 50 & 100 & 200 & 500 \\
    \cmidrule(rl){2-7}\cmidrule(rl){8-13}
    Base & 227 & 227 & 227 & 227 & 227 & 227 & 275 & 275 & 275 & 275 & 275 & 275 \\
    Lemmas & 226 & 242 & 246 & 258 & 257 & 258 & 278 & 285 & 284 & 281 & 283 & 284 \\
    \cmidrule(rl){2-7}\cmidrule(rl){8-13}
    Total & 231 & 243 & 247 & 258 & 257 & 258 & 282 & 285 & 284 & 283 & 284 & 285 \\
    \xbottomrule
  \end{tabular}  
\end{table}

\subsubsection{Linear vs GNN Model.}
The preceding experiments suggest that even a simple linear model can
provide useful guidance when features are carefully
selected. Table~\ref{tab:modeltype} shows that the GNN --- which
processes the formulas directly and has no access to expert designed
features --- also successfully learns to identify useful lemmas for \SGCD
and even slightly surpasses the linear model. \texttt{LCL125-1} can
only be solved by the GNN-assisted prover, even at extremely large
time limits.

\begin{table}[htb]
  \caption{Number of problems solved by SGCD over 2~iterations of training both a linear and a graph neural network model, for time limits 50~s, 100~s, 500~s and 30~min.}
  \label{tab:modeltype}
  \centering
  \begin{tabular}{ l @{\hspace{5pt}} r r r r @{\hspace{5pt}} r r r r}
    \xtoprule
    & \multicolumn{4}{c}{Linear} & \multicolumn{4}{c}{GNN} \\
    Time & 50s & 100s & 500s & 30m & 50s & 100s & 500s & 30m \\
    \cmidrule(rl){2-5}\cmidrule(rl){6-9}
    Base   & 266 & 275 & 285 & 285 & 266 & 275 & 285 & 285 \\
    Iter 1 & 280 & 282 & 284 & 281 & 272 & 282 & 283 & 284 \\
    Iter 2 & 281 & 283 & 281 & 283 & 279 & 282 & 282 & 284 \\
    \cmidrule(rl){2-5}\cmidrule(rl){6-9}
    Total  & 282 & 284 & 286 & 286 & 279 & 285 & 287 & 287 \\
    \xbottomrule
  \end{tabular}
\end{table}

\subsection{Discussion of Learning-Based Experiments}
When enhanced by learning-based lemma selection, \SGCD solves 287 of the 312
problems. These include 28 problems not solved by the leading first-order
prover \Vampire~\cite{vampire}, which solves 263 problems in its
\name{CASC}~\cite{CASC} portfolio mode. Supplemented with our lemmas, \Vampire
is boosted to 284 solved problems.
In combination, boosted \SGCD and \Vampire give 293 solved problems.
Taking into account the solutions obtained by further provers with our lemmas,
we obtain a total of 297. For detailed results see \appref{app:bestproofs}
and \url{http://cs.christophwernhard.com/cdtools/exp-lemmas/lemmas.html}.

A notable observation is that all systems --- with the exception of
\EProver\ --- improve when provided with selected lemmas. We argue that our
framework addresses fundamental weaknesses of both purely goal-driven systems
such as \CMProver, \leanCoP and \CCSV, as well as those of saturation style
systems such as \Vampire and \EProver. For the former, it is their inability
to generate lemmas, which results in unduly long proofs. For the latter, it is
their unrestricted expansion of the branching of the search space. We find
that goal-driven systems demonstrate huge improvement when lemmas are added:
usually 20--40\% depending on the configuration. The improvement is much more
modest for saturation style systems, partly because their baselines are
already stronger and partly because learned lemma selection still has a large
room for improvement. This is the focus of our immediate future work.
SGCD already provides a balance between goal-driven search and axiom-driven
lemma generation and we only see significant improvement from lemmas when the
time limit on proof search is smaller.
Our manual feature-based linear model allows for exploiting expert
knowledge. However, we see more potential in automated
feature extraction via GNNs. The fact that the two models perform similarly
suggests that we are not providing enough training data for the GNN to
manifest its full capabilities.

\section{Proving \texttt{LCL073-1}}
\label{sec:LCL073}

\texttt{LCL073-1} was proven by Meredith in the early 1950s with substitution
and detachment \cite{meredith:single:1953} but it remains outstandingly hard for
ATP, where it came to attention in 1992 \cite{mccune:wos:cd:1992};
TPTP reports rating 1.0 and status \name{Unknown} since 1997. Only Wos proved
it in the year 2000 with several invocations of \OTTER~\cite{wos:meredith},
transferring output and insight between runs.
The problem has a single axiom,
\[\P(\i(\i(\i(\i(\i(x,y),\i(\f{n}(z),\f{n}(u))),z),v),\i(\i(v,x),\i(u,x)))),\]
and the goal $\P(\i(\i(\fa,\fb),\i(\i(\fb,\fc),\i(\fa,\fc))))$, known as \Syll
\cite{ulrich:legacy:2001}.
The wider context is showing that a single axiom entails the elements of
a known axiomatization of a propositional logic.
Experiments with \SGCD in our workflow led to a proof of \texttt{LCL073-1}
(Fig.~\ref{fig-proof-lcl073-1}, also \appref{app:proof:lcl073-1})
surprisingly quickly. Its compacted size is 46, between that of Meredith (40,
reconstructed with CD in \cite{wos:meredith}) and that of Wos (74). Our
workflow is much simpler than Wos', basically the same as our other
experiments but restricted to one phase of lemma generation and incorporation,
with only heuristic lemma selection, no learning. Nevertheless, success is
fragile with respect to configuration, where reasons for failure or
success are not obvious.

\begin{figure}[t]
  \scriptsize
$2 = \D(1, \D(1, \D(1, 1))),\; 3 = \D(2, 2),\; 4 = \D(1, 3),\; 5 = \D(1, 4),\;
6 = \D(5, 1),\;
7 = \D(5, 6),\;\\
8 = \D(\D(\D(1, \D(1, 7)), 6), 1),\;
9 = \D(8, 6),\;
10 = \D(8, \D(1, 9)),\;
11 = \D(\D(1, \D(1, \D(4, 10))), 1),\\
12 = \D(1, \D(6, \D(1, \D(\D(1, \D(9, \D(9, \D(\D(11, 3), 4)))), 1)))),\;
13 = \D(\D(\D(12, \D(5, \D(8, 12))), 1), 7),\;\\
14 = \D(1, \D(13, \D(1, \D(13, 5)))),\;
15 = \D(\D(1, \D(13, \D(\D(\D(\D(13, 6), 9), 11), 10))), \D(14, \D(14, 1)))$
\caption{The D-term of our proof of \texttt{LCL073-1} represented by factor equations.}
  \label{fig-proof-lcl073-1}
\end{figure}

Our configuration parameters are not problem specific, although we started out
with lemma generation by PSP-level because it led earlier to a short proof
of \texttt{LCL038-1}~\cite{cw:sgcd,cwwb:article}.
We first call \SGCD to generate lemmas by PSP-level enumeration, configured with a cache
size of 5,000, terminating after 60~s with exhaustion of the search
space.\footnote{Notebook hardware, Intel\textsuperscript{\textregistered} Core\texttrademark{} i7-1260P processor, 32~GB RAM.}
Lemma features are computed for the 98,198 generated
lemmas and written to disk, taking another 120~s.
Lemmas are then ordered lexicographically according to five
features relating to sharing of symbols and subterms with the goal,
and to formula dimensions, taking a further ~70~s. These five features are
\pl{lf\_h\_height}, \pl{lf\_h\_excluded\_goal\_subterms}, \pl{lf\_h\_tsize},
\pl{lf\_h\_distinct\_vars}, \pl{dcterm\_hash}, see
\appref{app:manualfeatures} for their specification.
We now call \SGCD again, configured such that it
performs PSP-level enumeration for axiom-driven phases, interleaved with
level enumeration by height for goal-driven phases with $0$ as $\g{preAddMaxLevel}$.
It incorporates the first 2,900 ordered lemmas\footnote{2,900 is one of the
fragile parameters. Depending on features chosen for ordering lemmas,
there are ranges around 3,000 where the problem is solved.}
as input by \name{replacement} (Sect.~\ref{sec:background}).
The cache size limit is set to 1,500, a value used in other generally
successful configurations.
Formulas occurring as subformulas of an earlier-proven formula are excluded,
a variation of the \emph{organic} property~\cite{luk:tarski:aussagenkalkuel:1930,cwwb:article}.
The proof is then found in 20~s, total time elapsed about 270~s.

The D-term dimensions $\la \mathit{compacted\ size},
\mathit{tree\ size}, \mathit{height} \ra$ are $\la 46, 3276,
40\ra$, compared to Meredith's $\la 40, 6172, 30 \ra$\footnote{The
\emph{length} reported in \cite{wos:meredith} is the compacted size if also
the proofs of the two other goals required to prove completeness of the single
axiom are considered. The notion of compacted size straightforwardly
generalizes from trees to \emph{sets} of trees \cite{cwwb:article}.}
and Wos' $\la 74, 9207, 48\ra$.
The maximal size (occurrences of non-constant function symbols) of a lemma formula (MGT of a subproof) in the proof is~19,
the maximal height (tree height, disregarding the predicate symbol)~9, and the maximal number of variables~7.
Of the 46 lemmas in the proof 12 are present in the
2,900 input lemmas.
Among the 46 lemma formulas 35 are weakly organic \cite{cwwb:article} and
4 involve double negation.
N-simplification \cite{cwwb:article} applies to 65 occurrences but does not
effect a size reduction. The proof is S- and C-regular \cite{cwwb:article}.
Certain configurations of \SGCD for the proving phase also yield further
proofs. In experiments so far, these are enumerated after the presented
proof and have larger compacted size.

Proof structure enumeration by PSP-level \cite{cwwb:article} is the main key
to finding our proof of \texttt{LCL073-1}.
It is used for lemma generation and for axiom-driven proof search, whereas
goal-driven phases use height instead. The structure of
the proof reflects this: all steps with the exception of the root can be considered PSP steps,
i.e. one premise is a subproof of the other.
The particular challenge of the problem lies in the fact that it was solved by
a human (Meredith). Unlike in recent ATP successes for Boolos'
curious inference \cite{boolos:curious:1987,benzmueller:etal:boolos:2023},
where the key is two particular second-order lemmas, the key here is a
proof-structural \emph{principle} for building-up proofs by lemmas.
Intuitively it might express a form of economy, building proofs from proofs at
hand, that belonged to Meredith's repertoire.

\section{Conclusion}
\label{sec:conclusion}

We presented encouraging results about the use of lemmas in proof search.
Provers are provided with lemmas generated via structure enumeration, a
feature of the CM, and filtered with either learned guidance or manual
heuristics. As a first step with this new methodology, we focus on the
class of CD problems where we obtained
strong results with our own system and substantial improvement of
general first-order provers based on different paradigms, including the
long-time competition leader \Vampire. Moreover, our approach has led to the ---
in a sense first --- automatic proof for the well-known Meredith
single axiom problem with TPTP difficulty rating 1.0.

An important and novel aspect in our work was the explicit consideration of
proof structures, which for CD have a particularly simple form in D-terms. Proof
structures of the CM have a direct correspondence to these
\cite{cwwb:article}, such that the CM may guide the way to generalizations for
more expressive logics.
Another course of generalization is to move from unit lemmas, i.e. sharing of
\emph{subtrees} of D-terms, to more powerful lemmas. Preliminary work shows a
correspondence between Horn clause lemmas, D-terms with variables, proofs in
the connection structure calculus \cite{eder:cs:1989}, and combinatory
compression \cite{cw:ccs}.

The learning-based experiments show little difference in performance between
using a simple linear model and a more sophisticated graph neural network. We
believe this is due to the small problem corpus, which yields a limited
training signal. Hence, we plan to scale the system up to larger problem sets.

Our work also sheds new light on perspectives for the CM. It is well-known
that the lack of inherent lemma maintenance is a disadvantage of the CM
compared to resolution, which can be overcome with the connection structure
calculus~\cite{eder:cs:1989}, a generalization of the CM.
Here we see in experiments a drastic improvement of the CM-CT provers by supplementing their
input with externally generated lemmas.
\SGCD, which grew out of the CM-CT approach and integrates repeated
lemma generation into the proving process, keeps up with RS provers on CD problems,
and can even be applied to improve these by supplying its lemmas as additional input.

\subsubsection{Acknowledgments.} We thank Jens Otten for
inspiring discussions at the outset of the current project and anonymous
reviewers for helpful suggestions to improve the presentation.

\nocite{rezus:2020:witness} \nocite{luk:selected:1970}

\bibliographystyle{splncs04}
\bibliography{bibtableaux23}

\clearpage
\appendix

\renewcommand{\f}[1]{\texttt{#1}}
\newcommand{\m}[1]{\mathit{#1}}
\newcommand{\var}[1]{\textnormal{\textit{#1}}}
\newcommand{\lit}[1]{\textnormal{\textit{#1}}}
\newcommand{\propsig}[2]{\noindent \f{#1} : #2}
\newcommand{\propdesc}[1]{\par \hspace*{\fill}\begin{minipage}{0.9\textwidth}#1\end{minipage}\par\smallskip}
\section{Manually Extracted Lemma Features}
\label{app:manualfeatures}

Below we specify the features that are manually extracted from lemmas and used
our linear model (Sect.~\ref{sec:ml}). We also provide their types as follows:

\begin{itemize}
  \item \lit{NatNum} A natural number.
  \item \lit{NormalizedValue} A number between 0 and 1.
\end{itemize}

\subsection{Features of the Lemma's D-term}

\propsig{lf\_d\_csize}{\lit{NatNum}}\\
\propsig{lf\_d\_tsize}{\lit{NatNum}}\\
\propsig{lf\_d\_height}{\lit{NatNum}} \propdesc{Compacted size, tree size and
  height of the lemma's D-term.}

\propsig{lf\_d\_grd\_csize}{\lit{NatNum}} \propdesc{Compacted size of the
  lemma's D-term after replacing all variables with 0.}

\propsig{lf\_d\_major\_minor\_relation}{\lit{NatNum}}

\propdesc{Describes the structural relationship of the subproofs of the major
  and minor premise of the lemma. Can have the following values: 0 identical
  or D-term is atomic; 1 is a strict superterm; 2 is a strict subterm; 3 none
  of these relationships; 4 for nonground D-terms.}

\propsig{lf\_d\_number\_of\_terminals}{\lit{NatNum}}

\propdesc{Number of subterms in the lemma's D-term which are of the form
  \f{d($d_1$,$d_2$)} where neither of $d_1,d_2$ is a compound term.}

\subsection{Context Dependent Features of the Lemma's D-Term}
\label{sec-lfp-features}

The features depend on the lemma as embedded in a given proof. They are only
available for training data extracted from proofs.

\propsig{lfp\_containing\_proof}{\f{proof}}

\propdesc{The proof that contains the lemma. Specified as the atom that
  identifies the proof. See also \f{lf\_proof}.}

\propsig{lfp\_d\_occs}{\lit{NatNum}} \propdesc{If the lemma's D-term is ground,
  the number of occurrences of the lemma's D-term in the proof's D-term. If
  the lemma's D-term is non-ground, the number of subterm occurrences of the
  proof's D-term that are instances of the lemma's D-term (note that these
  subterm occurrences may overlap).}

\propsig{lfp\_d\_incoming}{\lit{NatNum}} \propdesc{The number of incoming edges
  of the lemma in the minimal DAG representing the proof's D-term. This can
  not be meaningfully determined for all lemma computation methods that yield
  s-lemmas.}

\propsig{lfp\_d\_occs\_innermost\_matches}{\lit{NatNum}} \propdesc{If the lemma's
  D-term is ground, the same as \f{lf\_d\_occs}. If the lemma's D-term has
  variables, the number of subterm occurrences of the proof's D-term that
  would be rewritten by \textsc{Innermost} replacement.}

\propsig{lfp\_d\_occs\_outermost\_matches}{\lit{NatNum}} \propdesc{If the lemma's
  D-term is ground, the same as \f{lf\_d\_occs}. If the lemma's D-term has
  variables, the number of subterm occurrences of the proof's D-term that
  would be rewritten by \textsc{Outermost} replacement.}

\propsig{lfp\_d\_min\_goal\_dist}{\lit{NatNum}} \propdesc{Number of edges in
  the proof's D-term of the shortest downward path from the root to a subtree
  that is an instance of the lemma's D-term.}

\subsection{Special Features of the Lemma's Formula Components}
  
\propsig{lf\_b\_length}{\lit{NatNum}}
\propdesc{Length of the \var{Body} component of the lemma's formula.}

\propsig{lf\_hb\_distinct\_hb\_shared\_vars}{\lit{NatNum}}
  \propdesc{Number of distinct
  variables that occur in the \var{Head} as well as in the \var{Body}
  component of the lemma's formula.}

\propsig{lf\_hb\_distinct\_h\_only\_vars}{\lit{NatNum}}
\propdesc{Number of distinct variables that occur in the \var{Head} but not
  the \var{Body} component of the lemma's formula.}

\propsig{lf\_hb\_distinct\_b\_only\_vars}{\lit{NatNum}}
\propdesc{Number of distinct variables that occur in the \var{Body} but not
  the \var{Head} component of the lemma's formula.}

\propsig{lf\_hb\_singletons}{\lit{NatNum}}
\propdesc{Number of distinct variables with a single occurrence in
  \var{Head} and \var{Body} taken together.}

\propsig{lf\_hb\_double\_negation\_occs}{\lit{NatNum}}
\propdesc{Number of instances of \f{n}(\f{n}(\_)) in \var{Head} and \var{Body}.}

\propsig{lf\_hb\_nongoal\_symbol\_occs}{\lit{NatNum}}
\propdesc{Number of occurrences of symbols (functions, constants) in
  \var{Head} and \var{Body} that do not appear in the problem's goal.
  Context-dependent in that it refers to the problem's goal formula.}

\propsig{lf\_h\_excluded\_goal\_subterms}{\lit{NatNum}}
\propdesc{Number of distinct subterms of the goal formula (after replacing
  constants systematicall by variables) that are not a subterm of \var{Head}
  (modulo the variant relationship). Context-dependent in that it refers to
  the problem's goal formula.}

\propsig{lf\_h\_subterms\_not\_in\_goal}{\lit{NatNum}}
\propdesc{Number of distinct subterms of the head that are not a subterm of
  the goal formula (after replacing constants systematicall by variables,
  modulo the variant relationship). Context-dependent in that it refers to the
  problem's goal formula.}

\propsig{lf\_hb\_compression\_ratio\_raw\_deflate}{\lit{NormalizedValue}}\\
\propsig{lf\_hb\_compression\_ratio\_treerepair}{\lit{NormalizedValue}}\\
\propsig{lf\_hb\_compression\_ratio\_dag}{\lit{NormalizedValue}}
\propdesc{Indicates how much the lemma's formula can be compressed. The value
  is roughly compressed size divided by original tree size. That is, formulas
  with ``much regularity'' such that they can be compressed stronger receive
  smaller values. The different properties realize this in variants for
  different notions and implementations of compression. The \f{raw\_deflate}
  version depends on intrinsics of SWI-Prolog's term representation and
  possibly gives different results for the same formula, depending on how it
  internally shares subterms.}

\propsig{lf\_hb\_organic}{\lit{NatNum}}

\propdesc{Whether the formula is organic. A nonenmpty \var{Body} is translated
  to an implication, e.g., if \var{Body} = [a,b,c], the considered formula is
  i(a,i(b,i(c,\var{Head}))). Determined with Minisat. Values: 0: the formula
  is organic; 1 the formula is not organic but weakly organic; 2 the formula
  is not weakly organic. See also
  \url{http://cs.christophwernhard.com/cdtools/downloads/cdtools/pldoc/organic_cd.html}.}

\propsig{lf\_hb\_name}{\lit{Atom}}

\propdesc{A name of the formula if it is well known under some name. For a
  formula with nonempty body the translation to implication is considered (as
  for \f{lf\_hb\_organic}) and the name is prefixed with \f{meta\_}. If the
  formula is not known under some name, the value is \f{zzz}. See also
  \f{named\_axiom}/2 in
  \url{http://cs.christophwernhard.com/cdtools/downloads/cdtools/pldoc/named_axioms_cd.html}.}

\propsig{lf\_hb\_name\_status}{\lit{NatNum}}

\propdesc{Number indicating whether the formula has a name in the sense of
  \f{lf\_hb\_name}: 0 if it has a name and an empty body, 1 if it has a
  nonempty body and a name (prefixed with \f{meta\_}), 2 otherwise.}

\subsection{General Features of the Lemma's Formula Components}

These features are specified below schematically with \var{COMP} for \f{h},
\f{b}, and \f{hb}, referring the respective features for the \var{Head}
component, the \var{Body} component and both of the components joined
together. The schema parameter \var{ITEM} for \f{var}, \f{const} and \f{fun}
refers to variables, constants (atomic values in Prolog syntax) and
function symbols with arity $\geq 1$, respectively.

\propsig{lf\_\var{COMP}\_csize}{\lit{NatNum}}\\
\propsig{lf\_\var{COMP}\_tsize}{\lit{NatNum}}\\
\propsig{lf\_\var{COMP}\_height}{\lit{NatNum}}

\propdesc{Compacted size, tree size and height, respectively. (We use these
  notions, which are also used for D-terms, here for formula terms.)}

\propsig{lf\_\var{COMP}\_distinct\_vars}{\lit{NatNum}}
\propdesc{Number of distinct variables.}

\propsig{lf\_\var{COMP}\_\var{ITEM}\_occs}{\lit{NatNum}}
\propdesc{Number of occurrences of syntactic objects of kind \var{ITEM}.}

\propsig{lf\_\var{COMP}\_occs\_of\_most\_frequent\_\var{ITEM}}{\lit{NatNum}}
\propdesc{Maximum number of occurrences of a syntactic object of kind \var{ITEM}.}

\clearpage
\section{Implementation Details of the Network Architecture}
\label{app:implementation}

This appendix supplements Sect.~\ref{sec:ml} with details of the 
network architecture and hardware setup.

\subsection*{Graph Neural Network Architecture}
We use a graph neural network with 8 convolution layers of 128
channels arranged into 4 residual blocks~\cite{residual-networks},
followed by a hidden dense layer of 1024 neurons and a final dense
layer that produces a single utility output.  Batch
normalization~\cite{batch-normalization} is applied before each
convolutional layer, and the non-linearity throughout is a rectified
linear unit~\cite{relu}.  The precise configuration was found by
manual optimization with respect to a holdout set.

\subsection*{Computation Used in the Experiments} We used a single
NVIDIA A100 GPU and 50 CPU cores (100 hyperthreads). Approximate times for a typical
experiment are: lemma extraction 20~min, a single iteration of proof
search 5~min, model training 50~min, lemma selection 60~min. So far,
we have run 151 experiments.
\clearpage
\section{Bar Chart Presentations of Selected Results}
\label{app:barchart}

For convenience of the reader, we provide the following bar chart
representations of selected data from Tables~\ref{tab:baseprovers},
\ref{tab:otherprovers} and~\ref{tab:lemmacount} in Sect.~\ref{sec:ml}.

\newlength{\prob}
\newcommand{\pbar}[2]{{\color{#1} \rule[-0.5pt]{#2\prob}{8.5pt}
    {\color{#1!30!black} #2}}}
\newcommand{\pbarpct}[2]{{\color{#1} \rule[-1.1pt]{#2\prob}{8.5pt} {\color{#1!60!black} #2\%}}}
\colorlet{pxa}{black!30!white}
\colorlet{pxb}{black!40!white}
\colorlet{pxc}{black!50!white}
\colorlet{pxd}{black}
\colorlet{pya}{black!20!white}
\colorlet{pyb}{black!25!white}
\colorlet{pyc}{black!30!white}
\colorlet{pyd}{black!35!white}
\colorlet{pye}{black!40!white}
\colorlet{pyf}{black!45!white}
\colorlet{pyg}{black!50!white}
\colorlet{pyh}{black}

\setlength{\prob}{0.862pt}

\begin{table}
  \caption{Performance of different provers over 2~iterations of
    training a linear model for 30~m time limit. Data from
    Table~\ref{tab:baseprovers}.}
  \begin{tabular}{L{8em}l}
  \SGCD Base & \pbar{pxa}{285}\\
  \SGCD Iter 1 & \pbar{pxb}{281}\\
  \SGCD Iter 2 & \pbar{pxc}{283}\\
  \SGCD Total & \pbar{pxd}{286}\\[5pt]   
  \ProverN Base & \pbar{pxa}{262}\\
  \ProverN Iter 1 & \pbar{pxb}{257}\\
  \ProverN Iter 2 & \pbar{pxc}{265}\\
  \ProverN Total & \pbar{pxd}{267}\\[5pt]    
  \CMProver Base & \pbar{pxa}{103}\\
  \CMProver Iter 1 & \pbar{pxb}{121}\\
  \CMProver Iter 2 & \pbar{pxc}{126}\\
  \CMProver Total & \pbar{pxd}{141}\\[5pt]
  \CCSV Base & \pbar{pxa}{105}\\
  \CCSV Iter 1 & \pbar{pxb}{130}\\
  \CCSV Iter 2 & \pbar{pxc}{128}\\
  \CCSV Total & \pbar{pxd}{145}\\    
\end{tabular}
\end{table}

\begin{table}[h!]
  \caption{Number of problems solved by \Vampire (casc), \EProver
    (autoschedule), \ProverN and \leanCoP without and with additional lemmas
    for 30~m time limit. Data from Table~\ref{tab:otherprovers}.}
  \begin{tabular}{L{8em}l}
    \Vampire Base & \pbar{pxa}{263}\\
    \Vampire Lemmas & \pbar{pxb}{283}\\
    \Vampire Total & \pbar{pxd}{284}\\[5pt]     
    \EProver Base & \pbar{pxa}{281}\\
    \EProver Lemmas & \pbar{pxb}{275}\\
    \EProver Total & \pbar{pxd}{286}\\[5pt]     
    \ProverN Base & \pbar{pxa}{260}\\
    \ProverN Lemmas & \pbar{pxb}{269}\\
    \ProverN Total & \pbar{pxd}{269}\\[5pt]     
    \leanCoP Base & \pbar{pxa}{77}\\
    \leanCoP Lemmas & \pbar{pxb}{113}\\
    \leanCoP Total & \pbar{pxd}{113}\\
  \end{tabular}
\end{table}

\clearpage

\begin{table}
  \centering
  \caption{Number of problems solved by \Vampire (casc) and \SGCD as we alter
    the number of supplemented lemmas between 10 and 500. We use a time limit
    of 100~s. Data from Table~\ref{tab:lemmacount}.}
  \begin{tabular}{lrl}
    Prover & \#Lemmas & \#Solved problems\\\midrule
    \Vampire & Base & \pbar{pya}{227}\\
    \Vampire & 10  & \pbar{pyb}{226}\\
    \Vampire & 25 & \pbar{pyc}{242}\\
    \Vampire & 50 & \pbar{pyd}{246}\\
    \Vampire & 100 & \pbar{pye}{258}\\
    \Vampire & 200 & \pbar{pyf}{257}\\
    \Vampire & 500 & \pbar{pyg}{258}\\[5pt]
    \SGCD & Base & \pbar{pya}{275}\\
    \SGCD & 10  & \pbar{pyb}{278}\\
    \SGCD & 25 & \pbar{pyc}{285}\\
    \SGCD & 50 & \pbar{pyd}{284}\\
    \SGCD & 100 & \pbar{pye}{281}\\
    \SGCD & 200 & \pbar{pyf}{283}\\
    \SGCD & 500 & \pbar{pyg}{284}\\

  \end{tabular}
\end{table}

\clearpage
\section{Lemma Usage}
\label{app:lemma:usage}

Table~\ref{tab-lemma-usage} shows for the experiments described in
Sect.~\ref{sec:ml} typical data on the usage of supplied input lemmas,
indicating how much proofs actually draw from the supplied lemmas. The
problems are either from the TPTP or, indicated by the \name{-tc} postfix in
the problem name (suggesting \name{Tarski's Claim} \cite{rezus:tc:2016}),
single-axiom problems derived from multi-axiom TPTP problems as described in
in Sect.~\ref{sec:experiment}. The data are from the first iteration of an
experiment with \SGCD, using a GNN without manual features and 30~min time
limit. Of the 312 problems, 60 where proven with lemmas. The remaining
problems were either proven already with lemma generation (224~problems) or
not proven (28~problems). The number of selected input lemmas was~200. Closure
under the subterm relationship for their D-terms led to 327--601 lemmas,
median~418 (column \textbf{A}). We refer to this set as the
\name{subproof-closed lemmas}. The compacted size of the 60~proofs was between
12 and 119, median 41.5 (column \textbf{B}). The ratio of compacted size, that
is, the cardinality of the set of distinct non-atomic subproofs to the
cardinality of the intersection of this set with the subproof-closed lemmas
(column \textbf{C}) was between 0.63 and 0.96, median 0.87 (column
\textbf{C}). This means that substantial portions of the proofs are actually
built-up from the supplied lemmas. If the subproof closure is not considered,
figures are quite different. The ratio with respect to the intersection with
the original set of 200~lemmas (column \textbf{E}) was then only between 0 and
0.48, median~0.14 (column~\textbf{F}).

\begin{table}
  \setlength{\tabcolsep}{3pt}
  \centering\scriptsize
  \caption{Usage of Learned Lemmas in Proofs}
  \label{tab-lemma-usage}
  \begin{tabular}{lrr@{\hspace{2em}}rr@{\hspace{2em}}rr}
    \textbf{Problem} & \textbf{A} & \textbf{B}
    & \textbf{C} & \textbf{D} & \textbf{E} & \textbf{F}\\\midrule
LCL019-1 & 445 & 51 & 45 & 0.88 & 3 & 0.06\\
LCL032-1 & 436 & 108 & 75 & 0.69 & 10 & 0.09\\
LCL037-1 & 463 & 119 & 75 & 0.63 & 8 & 0.07\\
LCL038-1 & 529 & 54 & 51 & 0.94 & 5 & 0.09\\
LCL054-1 & 504 & 54 & 50 & 0.93 & 4 & 0.07\\
LCL058-1 & 518 & 31 & 28 & 0.90 & 0 & 0.00\\
LCL060-1 & 460 & 40 & 36 & 0.90 & 8 & 0.20\\
LCL061-1 & 498 & 42 & 37 & 0.88 & 2 & 0.05\\
LCL062-1 & 493 & 45 & 38 & 0.84 & 6 & 0.13\\
LCL074-1 & 411 & 51 & 42 & 0.82 & 1 & 0.02\\
LCL084-2 & 520 & 53 & 50 & 0.94 & 6 & 0.11\\
LCL084-3 & 524 & 57 & 55 & 0.96 & 5 & 0.09\\
LCL100-1 & 508 & 23 & 18 & 0.78 & 0 & 0.00\\
LCL103-1 & 499 & 14 & 13 & 0.93 & 1 & 0.07\\
LCL122-1 & 510 & 35 & 32 & 0.91 & 3 & 0.09\\
LCL127-1 & 601 & 32 & 29 & 0.91 & 6 & 0.19\\
LCL129-1 & 491 & 12 & 11 & 0.92 & 2 & 0.17\\
LCL167-1 & 366 & 48 & 45 & 0.94 & 12 & 0.25\\
LCL374-1 & 482 & 42 & 38 & 0.90 & 7 & 0.17\\
LCL375-1 & 491 & 43 & 37 & 0.86 & 6 & 0.14\\
LCL376-1 & 504 & 30 & 26 & 0.87 & 2 & 0.07\\
LCL377-1 & 508 & 48 & 42 & 0.88 & 5 & 0.10\\
LCL388-1 & 500 & 54 & 51 & 0.94 & 2 & 0.04\\
LCL389-1 & 491 & 45 & 42 & 0.93 & 2 & 0.04\\
LCL391-1 & 455 & 70 & 62 & 0.89 & 18 & 0.26\\
LCL393-1 & 480 & 42 & 39 & 0.93 & 8 & 0.19\\
LCL394-1 & 480 & 44 & 39 & 0.89 & 7 & 0.16\\
LCL395-1 & 504 & 59 & 50 & 0.85 & 9 & 0.15\\
LCL403-1 & 483 & 59 & 56 & 0.95 & 7 & 0.12\\
LCL404-1 & 467 & 39 & 36 & 0.92 & 4 & 0.10\\
  \end{tabular}
  \hspace*{\fill}%
  \begin{tabular}{lrr@{\hspace{2em}}rr@{\hspace{2em}}rr}
    \textbf{Problem} & \textbf{A} & \textbf{B}
    & \textbf{C} & \textbf{D} & \textbf{E} & \textbf{F}\\\midrule
LCL028-1-tc & 407 & 35 & 29 & 0.83 & 3 & 0.09\\
LCL030-1-tc & 389 & 23 & 20 & 0.87 & 1 & 0.04\\
LCL031-1-tc & 388 & 35 & 31 & 0.89 & 2 & 0.06\\
LCL054-1-tc & 390 & 46 & 40 & 0.87 & 7 & 0.15\\
LCL058-1-tc & 399 & 34 & 31 & 0.91 & 5 & 0.15\\
LCL060-1-tc & 389 & 36 & 29 & 0.81 & 13 & 0.36\\
LCL061-1-tc & 368 & 37 & 28 & 0.76 & 9 & 0.24\\
LCL062-1-tc & 384 & 37 & 28 & 0.76 & 10 & 0.27\\
LCL084-2-tc & 331 & 46 & 32 & 0.70 & 19 & 0.41\\
LCL084-3-tc & 327 & 59 & 38 & 0.64 & 28 & 0.47\\
LCL114-1-tc & 425 & 42 & 39 & 0.93 & 2 & 0.05\\
LCL116-1-tc & 429 & 42 & 37 & 0.88 & 2 & 0.05\\
LCL373-1-tc & 368 & 47 & 40 & 0.85 & 10 & 0.21\\
LCL374-1-tc & 378 & 34 & 29 & 0.85 & 5 & 0.15\\
LCL375-1-tc & 384 & 35 & 26 & 0.74 & 5 & 0.14\\
LCL376-1-tc & 380 & 39 & 33 & 0.85 & 4 & 0.10\\
LCL377-1-tc & 386 & 41 & 34 & 0.83 & 6 & 0.15\\
LCL382-1-tc & 375 & 28 & 25 & 0.89 & 3 & 0.11\\
LCL383-1-tc & 394 & 32 & 29 & 0.91 & 3 & 0.09\\
LCL385-1-tc & 398 & 37 & 31 & 0.84 & 5 & 0.14\\
LCL388-1-tc & 402 & 37 & 31 & 0.84 & 6 & 0.16\\
LCL389-1-tc & 407 & 43 & 37 & 0.86 & 7 & 0.16\\
LCL390-1-tc & 384 & 42 & 35 & 0.83 & 9 & 0.21\\
LCL391-1-tc & 378 & 36 & 27 & 0.75 & 7 & 0.19\\
LCL392-1-tc & 384 & 25 & 22 & 0.88 & 6 & 0.24\\
LCL393-1-tc & 377 & 38 & 30 & 0.79 & 10 & 0.26\\
LCL394-1-tc & 384 & 39 & 30 & 0.77 & 11 & 0.28\\
LCL395-1-tc & 364 & 43 & 28 & 0.65 & 7 & 0.16\\
LCL403-1-tc & 365 & 34 & 28 & 0.82 & 10 & 0.29\\
LCL404-1-tc & 369 & 34 & 31 & 0.91 & 9 & 0.26\\
\end{tabular}
\end{table}

\clearpage
\section{''Best'' Encountered Proofs of Individual TPTP Problems}
\label{app:bestproofs}

\mathchardef\myhyphen="2D

Tables~\ref{tab-bestproofs-1} and~\ref{tab-bestproofs-2} below show for each
of the 196 pure CD problems in TPTP 8.1.2\footnote{Those CD problems that
remain after excluding from all 206~CD problems in the TPTP those two with
status \name{satisfiable}, those five with a form of detachment that is based
on implication represented by disjunction and negation, and those three with a
non-atomic goal theorem. In our experiments we also used further problems, 312
in total, derived from these TPTP problems.} characteristics of the ``best''
proof encountered in our experiments. Proofs were there ordered by lexical
comparison of the number tuple
\[\la \mathit{cdproof}, \mathit{csize}, \mathit{tsize}, \mathit{height}, \mathit{nsimp}, \mathit{lemma\myhyphen use},
\mathit{time} \ra.\] There $\mathit{cdproof}$ is $0$ or $1$ depending on
whether the proof is by CD ($0$) or some other calculus or unreported
$(1)$.\footnote{In applications with CD, e.g \cite{fitelson:walsh:2021},
usually CD proofs (see Sect.~\ref{sec:cd:atp}) are desired.} CD
proofs are obtained directly from \SGCD and \CCSV, and via straightforward
conversions from \ProverN and \CMProver. Compacted size, tree size and height
of the proof, after n-simplification \cite{cwwb:lukas:2021}, are considered as
$\mathit{csize}$, $\mathit{tsize}$ and $\mathit{height}$. If the proof is not
a CD proof, these values are not available. If n-simplification had no
reducing effect on these size measures, then $\mathit{nsimp}$ is $0$,
otherwise $1$. If the proof was obtained without input lemmas, then
$\mathit{lemma\myhyphen use}$ is $0$, otherwise~$1$. The time in seconds used
for proving (with input lemmas only for the last prover invocation with the
lemmas) is $\mathit{time}$.

The tables show for each problem the data of the ``best'' proof: The problem
and its rating in TPTP~8.1.2, the proof dimensions as
$\mathit{csize}/\mathit{tsize}/\mathit{height}$, prefixed with $n$ to indicate
that n-simplification had reducing effect, $\mathit{time}$ and the prover or
configuration that produced the proof. An asterisk (*) indicates that the
configuration involved lemma application. The suffix indicates the selection
method for the lemmas: -LIN* learning with a linear model, -GNN* learning with
a GNN model, -LIN-GNN* learning with a combined linear/GNN model, -HEU-X*
lemma selection by sorting according to heuristic features. For all shown
provers with exception of $\EProver$ and $\Vampire$, the -LIN*, -GNN* and
-LIN-GNN* configurations involved \emph{iterative improvement}. Also some
results for special configurations with heuristic lemma selection are
included, indicated by the following prover names: \ProverN-HEU-1* is Prover9
with 1,000 PSP optim input lemmas, \Vampire-HEU-2* is Vampire with 1,000 PSP
plain input lemmas and \SGCD-HEU-3* is the setup described in
Sect.~\ref{sec:LCL073}. The data from the unstarred versions are from base
runs of our experiments, before learning. A comprehensive table that extends
Tables~\ref{tab-bestproofs-1} and~\ref{tab-bestproofs-2} is at
\url{http://cs.christophwernhard.com/cdtools/exp-lemmas/lemmas.html}.

These tables for specific TPTP problems facilitate comparison with other
provers and approaches. They include solutions for 189 of the 196 problems,
where 6 of the 7 unsolved problems are rated 1.00 and one is rated 0.86. This
shows that our discourse indeed takes place at the very edge of the overall
state of the art of first-order proving, as far as CD problems are concerned.
The tables show that the lemma-enhanced configurations become more relevant
for problems with increased difficulty rating. However, for four of the most
difficult solved problems, lemmas selected with manually crafted heuristic led
to success. It remains to cover this also with machine-learned lemma
selection, where the identified difficult problems provide test cases. Only
four of the proven problems could be not be proven with CD proofs, two of them
actually quickly by \Vampire and \EProver, which calls for a deeper
investigation of their non-CD proofs, their possible proof translation to CD,
and the possible role of non-unit lemmas corresponding to binary resolution
there.

\begin{table}
  \centering
  \caption{Data of ``best'' encountered proofs I/II}
  \label{tab-bestproofs-1}
    \scriptsize
\scalebox{0.86}{
\begin{tabular}{lrrrl}
\textbf{Problem} & \textbf{Rtg} & \textbf{Dims} & \textbf{Time} & \textbf{Prover}\\\midrule
LCL006-1 & 0.00 & 5/7/4 & 0.06 & \CCSV\\
LCL007-1 & 0.00 & 1/1/1 & 0.01 & \SGCD\\
LCL008-1 & 0.00 & 5/5/5 & 0.01 & \SGCD\\
LCL009-1 & 0.00 & 7/17/6 & 0.06 & \CCSV\\
LCL010-1 & 0.00 & 5/8/5 & 0.03 & \SGCD\\
LCL011-1 & 0.00 & 7/16/7 & 0.07 & \CCSV\\
LCL013-1 & 0.00 & 2/2/2 & 0.01 & \SGCD\\
LCL015-1 & 0.00 & 24/73/19 & 162.96 & \SGCD\\
LCL022-1 & 0.00 & 8/33/7 & 0.46 & \CCSV\\
LCL023-1 & 0.00 & 7/18/7 & 0.09 & \CCSV\\
LCL025-1 & 0.00 & 6/9/6 & 0.23 & \CCSV\\
LCL026-1 & 0.00 & 22/29/15 & 2.76 & \SGCD\\
LCL027-1 & 0.00 & 3/3/3 & 0.01 & \SGCD\\
LCL029-1 & 0.00 & 7/14/6 & 10.79 & \CCSV\\
LCL033-1 & 0.00 & 6/7/6 & 0.02 & \SGCD\\
LCL034-1 & 0.00 & 24/46/19 & 1.85 & \SGCD\\
LCL035-1 & 0.00 & 5/6/5 & 0.02 & \SGCD\\
LCL036-1 & 0.00 & n 11/59/11 & 596.99 & \CCSV\\
LCL038-1 & 0.00 & 52/119/27 & 86.11 & \SGCD-LIN-GNN*\\
LCL041-1 & 0.00 & 3/3/2 & 0.03 & \SGCD\\
LCL043-1 & 0.00 & 2/2/2 & 0.01 & \SGCD\\
LCL044-1 & 0.00 & 3/3/3 & 0.02 & \SGCD\\
LCL045-1 & 0.00 & 5/5/4 & 0.06 & \CMProver\\
LCL046-1 & 0.00 & 2/2/2 & 0.01 & \SGCD\\
LCL047-1 & 0.00 & 16/22/5 & 0.62 & \SGCD\\
LCL048-1 & 0.00 & 16/19/13 & 113.13 & \SGCD\\
LCL049-1 & 0.00 & 20/24/14 & 105.82 & \SGCD\\
LCL050-1 & 0.00 & 22/27/17 & 112.76 & \SGCD\\
LCL051-1 & 0.00 & 21/26/10 & 128.74 & \SGCD\\
LCL052-1 & 0.00 & 17/26/13 & 69.77 & \SGCD\\
LCL053-1 & 0.00 & 18/28/15 & 81.14 & \SGCD\\
LCL055-1 & 0.00 & 15/24/11 & 50.73 & \SGCD\\
LCL056-1 & 0.00 & 16/25/12 & 76.68 & \SGCD\\
LCL057-1 & 0.00 & 20/30/16 & 241.43 & \SGCD\\
LCL058-1 & 0.00 & 30/68/17 & 1.15 & \SGCD-LIN*\\
LCL059-1 & 0.00 & 15/19/5 & 854.67 & \CMProver\\
LCL064-1 & 0.00 & 6/9/6 & 0.12 & \CCSV\\
LCL064-2 & 0.00 & 6/9/6 & 0.14 & \CCSV\\
LCL065-1 & 0.00 & 7/9/7 & 0.04 & \SGCD\\
LCL066-1 & 0.00 & 7/7/7 & 0.07 & \SGCD\\
LCL067-1 & 0.00 & 10/20/6 & 914.91 & \CCSV\\
LCL068-1 & 0.00 & 15/26/11 & 4.36 & \SGCD\\
LCL069-1 & 0.00 & 8/8/5 & 0.04 & \SGCD\\
LCL070-1 & 0.00 & 10/18/6 & 178.51 & \CCSV\\
LCL071-1 & 0.00 & 13/16/6 & 24.78 & \SGCD\\
LCL072-1 & 0.00 & 7/8/4 & 0.02 & \SGCD\\
LCL075-1 & 0.00 & 8/20/8 & 0.16 & \ProverN\\
LCL076-1 & 0.00 & 7/9/7 & 0.06 & \SGCD\\
LCL076-2 & 0.00 & 1/1/1 & 0.01 & \SGCD\\
\end{tabular}
\hspace{7pt}
\begin{tabular}{lrrrl}
\textbf{Problem} & \textbf{Rtg} & \textbf{Dims} & \textbf{Time} & \textbf{Prover}\\\midrule
LCL077-1 & 0.00 & 6/8/6 & 0.05 & \SGCD\\
LCL079-1 & 0.00 & 3/3/3 & 0.02 & \SGCD\\
LCL080-1 & 0.00 & 9/10/8 & 383.43 & \CCSV\\
LCL080-2 & 0.00 & 9/12/6 & 0.31 & \SGCD\\
LCL081-1 & 0.00 & 6/10/6 & 0.06 & \SGCD\\
LCL082-1 & 0.00 & 6/7/6 & 0.02 & \SGCD\\
LCL083-1 & 0.00 & 11/15/11 & 0.35 & \SGCD\\
LCL083-2 & 0.00 & 8/9/8 & 0.06 & \SGCD\\
LCL085-1 & 0.00 & 23/29/20 & 13.85 & \SGCD\\
LCL086-1 & 0.00 & n 10/22/8 & 77.91 & \CCSV\\
LCL087-1 & 0.00 & 8/12/8 & 0.03 & \SGCD\\
LCL088-1 & 0.00 & 12/18/8 & 1157.65 & \CMProver\\
LCL089-1 & 0.00 & n 10/27/10 & 254.25 & \CCSV\\
LCL090-1 & 0.00 & 14/20/14 & 203.68 & \SGCD\\
LCL091-1 & 0.00 & 11/16/10 & 35.83 & \CMProver\\
LCL092-1 & 0.00 & n 12/34/11 & 369.00 & \CCSV\\
LCL093-1 & 0.00 & 25/29/16 & 103.38 & \SGCD\\
LCL094-1 & 0.00 & 16/26/11 & 0.56 & \SGCD\\
LCL095-1 & 0.00 & 17/21/16 & 143.91 & \SGCD\\
LCL096-1 & 0.00 & 4/4/3 & 0.02 & \SGCD\\
LCL097-1 & 0.00 & 4/6/4 & 0.03 & \SGCD\\
LCL098-1 & 0.00 & 4/6/4 & 0.02 & \SGCD\\
LCL101-1 & 0.00 & 7/14/6 & 0.06 & \CCSV\\
LCL102-1 & 0.00 & 7/7/4 & 0.06 & \CMProver\\
LCL103-1 & 0.00 & 10/16/9 & 984.38 & \CCSV\\
LCL104-1 & 0.00 & 6/15/5 & 0.09 & \CCSV\\
LCL106-1 & 0.00 & 4/4/4 & 0.01 & \SGCD\\
LCL107-1 & 0.00 & 5/9/5 & 0.03 & \SGCD\\
LCL108-1 & 0.00 & 7/20/6 & 0.04 & \ProverN\\
LCL110-1 & 0.00 & 7/9/6 & 1.01 & \CCSV\\
LCL111-1 & 0.00 & 5/5/3 & 0.03 & \SGCD\\
LCL112-1 & 0.00 & 8/10/7 & 6.77 & \CCSV\\
LCL113-1 & 0.00 & 15/18/5 & 1286.31 & \CMProver\\
LCL114-1 & 0.00 & 21/31/8 & 525.31 & \SGCD\\
LCL115-1 & 0.00 & 11/16/9 & 0.24 & \SGCD\\
LCL116-1 & 0.00 & 24/52/13 & 10.65 & \ProverN\\
LCL117-1 & 0.00 & 3/4/3 & 0.03 & \SGCD\\
LCL118-1 & 0.00 & 7/8/7 & 0.02 & \SGCD\\
LCL120-1 & 0.00 & 6/7/6 & 0.03 & \SGCD\\
LCL121-1 & 0.00 & 12/92/12 & 649.07 & \CCSV\\
LCL123-1 & 0.00 & 10/45/7 & 0.75 & \CCSV\\
LCL126-1 & 0.00 & 4/4/4 & 0.01 & \SGCD\\
LCL127-1 & 0.00 & 22/49/13 & 116.04 & \CMProver-LIN*\\
LCL128-1 & 0.00 & 24/352/19 & 4.05 & \ProverN\\
LCL129-1 & 0.00 & 11/42/11 & 336.22 & \CCSV\\
LCL130-1 & 0.00 & 5/8/5 & 0.02 & \SGCD\\
LCL131-1 & 0.00 & 11/50/11 & 125.01 & \CCSV\\
LCL256-1 & 0.00 & 20/31/13 & 114.94 & \SGCD\\
LCL257-1 & 0.00 & 7/13/6 & 0.09 & \CCSV\\
\end{tabular}}
\end{table}

\clearpage

\begin{table}
  \centering
  \caption{Data of ``best'' encountered proofs II/II}  
  \label{tab-bestproofs-2}
    \scriptsize
\scalebox{0.86}{
    \begin{tabular}{lrrrl}
\textbf{Problem} & \textbf{Rtg} & \textbf{Dims} & \textbf{Time} & \textbf{Prover}\\\midrule  
LCL355-1 & 0.00 & 1/1/1 & 0.01 & \SGCD\\
LCL356-1 & 0.00 & 2/3/2 & 0.01 & \SGCD\\
LCL357-1 & 0.00 & 2/2/2 & 0.01 & \SGCD\\
LCL358-1 & 0.00 & 4/5/4 & 0.02 & \SGCD\\
LCL359-1 & 0.00 & 3/5/3 & 0.02 & \SGCD\\
LCL360-1 & 0.00 & 1/1/1 & 0.01 & \SGCD\\
LCL361-1 & 0.00 & 4/4/3 & 0.02 & \SGCD\\
LCL362-1 & 0.00 & 4/4/4 & 0.02 & \SGCD\\
LCL363-1 & 0.00 & 6/6/5 & 0.03 & \SGCD\\
LCL364-1 & 0.00 & 9/14/8 & 45.56 & \CCSV\\
LCL366-1 & 0.00 & 14/16/5 & 693.32 & \CMProver\\
LCL367-1 & 0.00 & 15/19/5 & 1.42 & \SGCD\\
LCL378-1 & 0.00 & 14/23/10 & 38.98 & \SGCD\\
LCL379-1 & 0.00 & 19/28/15 & 99.51 & \SGCD\\
LCL380-1 & 0.00 & 17/26/13 & 225.93 & \SGCD\\
LCL381-1 & 0.00 & 18/27/14 & 4.67 & \SGCD\\
LCL385-1 & 0.00 & 30/44/16 & 293.39 & \SGCD\\
LCL386-1 & 0.00 & 27/39/14 & 234.04 & \SGCD\\
LCL387-1 & 0.00 & 27/46/14 & 15.98 & \SGCD\\
LCL388-1 & 0.00 & 38/109/19 & 92.99 & \SGCD\\
LCL389-1 & 0.00 & 37/411/22 & 4.69 & \ProverN\\
LCL396-1 & 0.00 & 21/31/17 & 193.84 & \SGCD\\
LCL397-1 & 0.00 & 7/8/7 & 0.17 & \SGCD\\
LCL398-1 & 0.00 & 3/3/3 & 0.04 & \SGCD\\
LCL399-1 & 0.00 & 12/13/11 & 6.45 & \SGCD\\
LCL400-1 & 0.00 & n 31/233/20 & 0.85 & \ProverN\\
LCL401-1 & 0.00 & 29/62/15 & 320.04 & \SGCD\\
LCL402-1 & 0.00 & n 30/191/20 & 0.41 & \ProverN\\
LCL404-1 & 0.00 & 36/409/22 & 4.36 & \ProverN\\
LCL405-1 & 0.00 & 26/34/14 & 199.42 & \SGCD\\
LCL416-1 & 0.00 & 10/17/8 & 14.75 & \SGCD\\
LCL012-1 & 0.14 & 18/31/16 & 62.51 & \SGCD\\
LCL014-1 & 0.14 & 10/15/7 & 11.58 & \CCSV\\
LCL016-1 & 0.14 & 39/89/14 & 195.72 & \SGCD\\
LCL017-1 & 0.14 & 50/129/17 & 194.28 & \SGCD\\
LCL018-1 & 0.14 & 22/54/16 & 147.63 & \SGCD\\
LCL019-1 & 0.14 & 35/132/18 & 202.29 & \SGCD\\
LCL021-1 & 0.14 & 68/225/19 & 331.46 & \SGCD\\
LCL024-1 & 0.14 & 10/15/9 & 79.22 & \CMProver\\
LCL030-1 & 0.14 & 8/16/6 & 62.68 & \CCSV\\
LCL031-1 & 0.14 & 23/26/10 & 209.34 & \SGCD\\
LCL040-1 & 0.14 & 8/9/5 & 15.85 & \CCSV\\
LCL042-1 & 0.14 & 8/8/4 & 499.52 & \CCSV\\
LCL054-1 & 0.14 & 33/283/21 & 218.11 & \ProverN\\
LCL060-1 & 0.14 & 36/76/17 & 0.41 & \CCSV-LIN*\\
LCL084-2 & 0.14 & n 53/87/25 & 80.13 & \SGCD-GNN*\\
LCL084-3 & 0.14 & 52/98/23 & 295.65 & \SGCD-LIN-GNN*\\
LCL100-1 & 0.14 & 18/44/11 & 187.43 & \CCSV-LIN*\\
LCL122-1 & 0.14 & 25/142/10 & 13.46 & \SGCD-LIN*\\
\end{tabular}
\hspace{7pt}
\begin{tabular}{lrrrl}  
\textbf{Problem} & \textbf{Rtg} & \textbf{Dims} & \textbf{Time} & \textbf{Prover}\\\midrule  
LCL166-1 & 0.14 & 51/155/18 & 182.67 & \SGCD\\
LCL369-1 & 0.14 & 22/34/16 & 2.82 & \SGCD\\
LCL370-1 & 0.14 & 27/39/13 & 242.93 & \SGCD\\
LCL371-1 & 0.14 & 27/39/13 & 233.94 & \SGCD\\
LCL373-1 & 0.14 & 37/68/14 & 708.81 & \SGCD\\
LCL382-1 & 0.14 & 29/53/18 & 6.21 & \SGCD\\
LCL384-1 & 0.14 & 13/23/5 & 683.01 & \CMProver\\
LCL390-1 & 0.14 & 31/45/14 & 281.13 & \SGCD\\
LCL403-1 & 0.14 & 40/94/16 & 30.54 & \SGCD-LIN*\\
LCL032-1 & 0.29 & n 67/15362/35 & 106.73 & \ProverN\\
LCL099-1 & 0.29 & 20/41/6 & 459.30 & \SGCD\\
LCL105-1 & 0.29 & 37/109/11 & 90.54 & \ProverN-LIN*\\
LCL119-1 & 0.29 & 83/28624/27 & 76.07 & \ProverN\\
LCL365-1 & 0.29 & 10/15/9 & 429.17 & \CCSV\\
LCL368-1 & 0.29 & 21/32/16 & 2.10 & \SGCD\\
LCL372-1 & 0.29 & 27/46/13 & 12.87 & \SGCD\\
LCL383-1 & 0.29 & 33/52/15 & 41.99 & \SGCD\\
LCL391-1 & 0.29 & 40/161/20 & 65.93 & \SGCD\\
LCL392-1 & 0.29 & 30/52/14 & 26.83 & \SGCD\\
LCL393-1 & 0.29 & 37/87/17 & 46.13 & \SGCD\\
LCL020-1 & 0.43 & 106/24989/37 & 21.65 & \ProverN-LIN*\\
LCL028-1 & 0.43 & 34/67/15 & 295.28 & \SGCD\\
LCL061-1 & 0.43 & 39/92/16 & 87.96 & \SGCD\\
LCL062-1 & 0.43 & 44/115/21 & 285.10 & \SGCD-LIN*\\
LCL124-1 & 0.43 & 27/130/10 & 76.25 & \SGCD-LIN*\\
LCL125-1 & 0.43 & 33/460/16 & 33.14 & \ProverN\\
LCL167-1 & 0.43 & 48/265/22 & 27.53 & \SGCD-GNN*\\
LCL374-1 & 0.43 & 33/77/17 & 42.47 & \SGCD\\
LCL375-1 & 0.43 & 43/103/20 & 56.44 & \SGCD-LIN*\\
LCL376-1 & 0.43 & 30/76/15 & 58.17 & \SGCD-GNN*\\
LCL394-1 & 0.43 & 41/81/17 & 267.22 & \SGCD\\
LCL875-1 & 0.43 & & 0.298 & \Vampire\\
LCL037-1 & 0.57 & n 72/45359/39 & 172.29 & \ProverN\\
LCL074-1 & 0.57 & n 50/136/18 & 998.93 & \SGCD\\
LCL377-1 & 0.57 & 38/78/15 & 62.71 & \SGCD\\
LCL395-1 & 0.57 & 45/112/20 & 140.94 & \SGCD\\
LCL428-1 & 0.57 & & 0.227 & \EProver\\
LCL109-1 & 0.86 & 72/348/22 & 226.55 & \ProverN-HEU-1*\\
LCL417-1 & 0.86 & & 647.386 & \Vampire-HEU-2*\\
LCL422-1 & 0.86 & & &\\
LCL876+1 & 0.93 & 70/396/22 & 227.17 & \ProverN-HEU-1*\\
LCL063-1 & 1.00 & & 943.481 & \EProver\\
LCL073-1 & 1.00 & 46/3276/40 & 16.55 & \SGCD-HEU-3*\\
LCL418-1 & 1.00 & & &\\
LCL419-1 & 1.00 & & &\\
LCL420-1 & 1.00 & & &\\
LCL421-1 & 1.00 & & &\\
LCL425-1 & 1.00 & & &\\
LCL426-1 & 1.00 & & &\\
\end{tabular}}
\end{table}

\clearpage
\newcommand{\mereq}{\hspace{0.15em}=\hspace{0.15em}}

\section{Our Proof of \texttt{LCL073-1} with MGTs}
\label{app:proof:lcl073-1}

The following figures present our proof of \texttt{LCL073-1}
(Sect.~\ref{sec:LCL073}) together with its lemma formulas, the MGTs of
subproofs, in the notation for CD applications used in the classical
literature.

\begin{figure}[h]
\begin{tabular}{rl}
1. & $\g{CCCCCpqCNrNsrtCCtpCsp}$\\
2. & $\g{CCCppqCrq} \mereq \D1 \D1 \D1 1$\\
3. & $\g{CpCqCrr} \mereq \D2 2$\\
4. & $\g{CCCpCqqrCsr} \mereq \D1 3$\\
5. & $\g{CCCpqrCqr} \mereq \D1 4$\\
6. & $\g{CpCCpqCrq} \mereq \D5 1 $\\
7. & $\g{CpCCCqprCsr} \mereq \D5 6$\\
8. & $\g{CCCpqrCCCqsCNrNpr} \mereq \D\D\D1 \D1 7 6 \mathrm{n}$\\
9. & $\g{CCCpqCNCCCrpsCtsNrCCCrpsCts} \mereq \D8 6$\\
10. & $\g{CCCpqCNCrpNCCCsCptuCvuCrp} \mereq \D8 \D1 9$\\
11. & $\g{CpCCqrCCNqNpr} \mereq  \D\D1\D1\D4.10.1$\\
12. & $\g{CCCpNqCCqrCsrCtCCqrCsr} \mereq
\D1\D6\D1\D\D1\D9\D9\D\D11.341$\\
13. & $\g{CCpCqrCCCsCtNprCqr} \mereq \D\D\D12.\D5\D8.12.\mathrm{n} 7$\\
14. & $\g{CCCCCpqrstCCCqrst} \mereq \D1\D13.\D1\D13.5$\\
*15. & $\g{CCpqCCqrCpr} \mereq
\D\D1\D13.\D\D\D\D13.69.11.10.\D14.\D14.1$\\
\end{tabular}

\caption{Our proof of \texttt{LCL073-1} in the notation of Meredith
  \cite{meredith:notes:1963,cwwb:article}. For each factor of the overall
  D-term the argument term of its MGT is there shown in \Lukasiewicz's
  notation.}
\end{figure}
\begin{figure}[h!]
\begin{tabular}{rl}
1. & $\P(\i(\i(\i(\i(\i(p, q), \i(\f{n}(r), \f{n}(s))), r), t), \i(\i(t, p), \i(s, p))))$\\
2. & $\P(\i(\i(\i(p,p),q),\i(r,q))) \mereq \D(1, \D(1, \D(1, 1)))$\\
3. & $\P(\i(p,\i(q,\i(r,r)))) \mereq \D(2, 2)$\\
4. & $\P(\i(\i(\i(p,\i(q,q)),r),\i(s,r))) \mereq \D(1, 3)$\\
5. & $\P(\i(\i(\i(p,q),r),\i(q,r))) \mereq \D(1, 4)$\\
6. & $\P(\i(p,\i(\i(p,q),\i(r,q)))) \mereq \D(5, 1) $\\
7. & $\P(\i(p,\i(\i(\i(q,p),r),\i(s,r)))) \mereq \D(5, 6)$\\
8. & $\P(\i(\i(\i(p,q),r),\i(\i(\i(q,s),\i(\f{n}(r),\f{n}(p))),r))) \mereq \D(\D(\D(1, \D(1, 7)), 6), \mathrm{n})$\\
9. & $\P(\i(\i(\i(p,q),\i(\f{n}(\i(\i(\i(r,p),s),\i(t,s))),\f{n}(r))),\i(\i(\i(r,p),s),\i(t,s)))) \mereq \D(8, 6)$\\
10. & $\P(\i(\i(\i(p,q),\i(\f{n}(\i(r,p)),\f{n}(\i(\i(\i(s,\i(p,t)),u),\i(v,u))))),\i(r,p))) \mereq \D(8, \D(1, 9))$\\
11. & $\P(\i(p,\i(\i(q,r),\i(\i(\f{n}(q),\f{n}(p)),r)))) \mereq  \D(\D(1, \D(1, \D(4, 10))), 1)$\\
12. & $\P(\i(\i(\i(p,\f{n}(q)),\i(\i(q,r),\i(s,r))),\i(t,\i(\i(q,r),\i(s,r)))))$\\& $\mereq\;
\D(1, \D(6, \D(1, \D(\D(1, \D(9, \D(9, \D(\D(11, 3), 4)))), 1))))$\\
13. & $\P(\i(\i(p,\i(q,r)),\i(\i(\i(s,\i(t,\f{n}(p))),r),\i(q,r)))) \mereq
\D(\D(\D(12, \D(5, \D(8, 12))), \mathrm{n}), 7)$\\
14. & $\P(\i(\i(\i(\i(\i(p,q),r),s),t),\i(\i(\i(q,r),s),t))) \mereq \D(1, \D(13, \D(1, \D(13, 5))))$\\
*15. & $\P(\i(\i(p,q),\i(\i(q,r),\i(p,r))))$\\& $\mereq\;
\D(\D(1, \D(13, \D(\D(\D(\D(13, 6), 9), 11), 10))), \D(14, \D(14,1)))$\\
\end{tabular}

\caption{Our proof of \texttt{LCL073-1} with the ATP-oriented notation for
  formulas and D-terms of the paper, arranged such that it mimics Meredith's
  notation.}
\end{figure}

\end{document}